\documentclass[fleqn,usenatbib]{mnras}

\usepackage{newtxtext,newtxmath}

\usepackage[T1]{fontenc}
\usepackage{pdflscape}

\DeclareRobustCommand{\VAN}[3]{#2}
\let\VANthebibliography\thebibliography
\def\thebibliography{\DeclareRobustCommand{\VAN}[3]{##3}\VANthebibliography}

\usepackage{graphicx}	
\usepackage{amsmath}	

\title[Cyclotron line in 1A 0535+262]{Cyclotron line evolution revealed with pulse-to-pulse analysis in the 2020 outburst of 1A 0535+262}

\author[Qing C. Shui et al.]{Qing C. Shui,$^{1,2}$\thanks{E-mail: shuiqc@ihep.ac.cn (IHEP)}
S. Zhang,$^{1}$\thanks{E-mail: szhang@ihep.ac.cn (IHEP)}
Peng J. Wang,$^{3}$\thanks{E-mail: wangpj@ihep.ac.cn (IHEP)}
Alexander A. Mushtukov,$^{4,5}$
A. Santangelo,$^{3}$
\newauthor
Shuang N. Zhang,$^{1,2}$
Ling D. Kong,$^{3}$
L. Ji,$^{6}$
Yu P. Chen,$^{1}$
V. Doroshenko,$^{3,7}$
F. Frontera,$^{8}$
Z. Chang,$^{1}$
\newauthor
Jing Q. Peng,$^{1,2}$
Hong X. Yin,$^{9}$
Jin L. Qu,$^{1}$
L. Tao,$^{1}$
Ming Y. Ge,$^{1}$
J. Li,$^{10,11}$
Wen T. Ye,$^{1,2}$
and Pan P. Li$^{1,2}$
\\
$^{1}$Key Laboratory of Particle Astrophysics, Institute of High Energy Physics, Chinese Academy of Sciences, 100049, Beijing, China\\
$^{2}$University of Chinese Academy of Sciences, Chinese Academy of Sciences, 100049, Beijing, China\\
$^{3}$Institut f\"{u}r Astronomie und Astrophysik, Kepler Center for Astro and Particle Physics, Eberhard Karls, Universit\"{a}t, Sand 1, D-72076 T\"{u}bingen, Germany\\
$^{4}$Astrophysics, Department of Physics, University of Oxford, Denys Wilkinson Building, Keble Road, Oxford OX1 3RH, UK\\
$^{5}$Leiden Observatory, Leiden University, NL-2300 RA Leiden, the Netherlands\\
$^{6}$School of Physics and Astronomy, Sun Yat-Sen University, Zhuhai, 519082, China\\
$^{7}$Space Research Institute of the Russian Academy of Sciences, Profsoyuznaya Str 84/32, Moscow 117997, Russia\\
$^{8}$Department of Physics and Earth Science, University of Ferrara, Via Saragat, 1 I-44122 Ferrara, Italy\\
$^{9}$Shandong Key Laboratory of Optical Astronomy and Solar-Terrestrial Environment, School of Space Science and Physics, Institute of Space Sciences,\\ \ Shandong University, Weihai, Shandong 264209, China\\
$^{10}$CAS Key Laboratory for Research in Galaxies and Cosmology, Department of Astronomy, University of Science and Technology of China, Hefei 230026, China\\
$^{11}$School of Astronomy and Space Science, University of Science and Technology of China, Hefei 230026, China
}

\date{Accepted XXX. Received YYY; in original form ZZZ}

\pubyear{2022}

\begin{document}
\label{firstpage}
\pagerange{\pageref{firstpage}--\pageref{lastpage}}
\maketitle

\begin{abstract}
We present a detailed analysis of the X-ray luminosity ($L_{\rm X}$) dependence of the cyclotron absorption line energy ($E_{\rm cyc}$) for the X-ray binary pulsar 1A 0535+262 during its 2020 giant outburst based on pulse-to-pulse analysis. By applying this technique to high cadence observations of Insight-HXMT, we reveal the most comprehensive $E_{\rm cyc}$-$L_{\rm X}$ correlation across a broad luminosity range of $\sim(0.03$--$1.3)\times10^{38}\ {\rm erg\ s^{-1}}$. Apart from the positive and negative correlations between cyclotron line energy and luminosity at $L_{\rm X}\sim(1$--$3)\times10^{37}\ {\rm erg\ s ^{-1}}$ and $\sim(7$--$13)\times10^{37}\ {\rm erg\ s ^{-1}}$, which are expected from the typical subcritical and supercritical accretion regimes, respectively, a plateau in the correlation is also detected at $\sim(3$--$7)\times10^{37}\ {\rm erg\ s^{-1}}$. 
Moreover, at the lowest luminosity level ($L_{\rm X}\lesssim10^{37}\ {\rm erg\ s^{-1}}$), the positive $E_{\rm cyc}$-$L_{\rm X}$ correlation seems to be broken, and the pulse profile also occurs a significant transition. These discoveries provide the first complete view on the correlation between luminosity and the centriod energy of the cyclotron line, and therefore are relevant for understanding how accretion onto magnetized neutron stars depends on luminosity.
\end{abstract}
\begin{keywords}
accretion, accretion discs -- X-ray: stars -- pulasrs: individual: 1A 0535+262
\end{keywords}



\section{Introduction} \label{sec:intro}
X-ray pulsars (XRPs) are strongly magnetized rotating neutron stars (NSs) in binary systems, accreting matter from their massive (spectral type late O or early B) companion star \citep[see][for reviews]{2015A&ARv..23....2W,2023hxga.book..138M,2023A&A...677A.134N}. Because of the strong magnetic field ($B\gtrsim10^{12}$ G), magnetic pressure dominates the ram pressure of the accreting gas within the relatively large Alfv\'{e}n radius of $\sim10^{7-10}$ cm. In this case, the fully-ionized accreting plasma is funnelled by the magnetic field  lines onto the magnetic poles of the NS, in which the kinetic energy of the in-falling matter is converted to heat and radiation. 
A strong magnetic field in the vicinity of the NS surface could quantize the energy of electrons according to the Landau levels \citep[see e.g.][]{1998ApJ...505..688I,2000ApJ...544.1067A}. Resonant scattering of photons in these electrons then leads to the generation of resonance features (generally appearing in absorption) in X-ray spectra of XRPs. These absorption-like features are called cyclotron resonant scattering features (CRSFs), and also called cyclotron lines \citep[see][for a recent review]{2019A&A...622A..61S}. The centroid energy of the fundamental cyclotron line, $E_{\rm cyc}$, is directly related to the magnetic field strength of the line-forming regions as $E_{\rm cyc}\approx11.6\times B_{12}$ keV, where $B_{12}$ is the magnetic field strength in units of $10^{12}$ G, hence permits a direct measurement of the magnetic field strength near the NS surface.

Luminosity ($L_{\rm X}$) dependence of the X-ray spectrum has been found in a number of XRPs \citep[see e.g.][]{1998AdSpR..22..987M,2013A&A...551A...1R}. In particular, two different types of luminosity dependence of the cyclotron line energy have been detected \citep[][]{2019A&A...622A..61S}. So far, a positive correlation between $E_{\rm cyc}$ and $L_{\rm X}$ has been observed in several sources, e.g. Her~X-1 \citep[][]{2007A&A...465L..25S}, Vela~X-1 \citep[][]{2014ApJ...780..133F,2016MNRAS.463..185L}, GX 304-1 \citep[][]{2011PASJ...63S.751Y,2015A&A...581A.121M,2017MNRAS.466.2752R} and Cep~X-4 \citep[][]{2017A&A...601A.126V}, while a clear $E_{\rm cyc}$-$L_{\rm X}$ anti-correlation has been only found in V~0332+53 \citep[][]{1990ApJ...365L..59M,2006MNRAS.371...19T,2010MNRAS.401.1628T,2016MNRAS.460L..99C,2017MNRAS.466.2143D,2018A&A...610A..88V} and 1A~0535+262 \citep[][]{2021ApJ...917L..38K}. In these two sources, a transition for correlation between positive and negative has been observed as well. In general, the negative $E_{\rm cyc}$-$L_{\rm X}$ correlation is detected in relatively high luminosity sources (e.g. V~0332+53), while the positive one is seen in lower luminosity sources (e.g. Her~X-1). 
Accordingly, the observed bimodal behavior should be related to the mass accretion rate, which determines the structure of the accretion flow near the NS surface. At high mass accretion rates, 
the radiative pressure is strong enough to stop the falling matter near the polar cap, and hence supports an extended emission region above the NS surface, known as the accretion column \citep[][]{1976MNRAS.175..395B}. In this case, the accretion column grows higher with the increasing accretion rate. At low mass accretion rates, the gas deceleration is presumably dominated by Coulomb interactions \citep[][]{1991ApJ...367..575B,2007A&A...465L..25S,2012A&A...544A.123B}, hence the characteristic emission height decreases with the increasing accretion rate. The transition between the aforementioned two different accretion regimes occurs at the so-called critical luminosity, $L_{\rm crit}$, which is associated with the onset of the accretion column \citep[][]{1976MNRAS.175..395B,2012A&A...544A.123B,2015MNRAS.447.1847M}. Since the cyclotron line energy is expected to reflect the characteristic emission height \citep[][]{2012A&A...544A.123B} and/or the bulk velocity of in-falling plasma \citep[][]{2015MNRAS.454.2714M}, the $E_{\rm cyc}$-$L_{\rm X}$ relation provides an almost unambiguous technique to trace the different accretion regimes at work at the magnetic pole of an accreting neutron star \citep[see e.g.][]{2017MNRAS.466.2143D,2018A&A...610A..88V,2021ApJ...917L..38K}. 

In addition to the spectral properties, the angular pattern of the emitted radiation also depends on the mass accretion rate, resulting in the luminosity dependence of the pulse profile \citep[][]{1976MNRAS.175..395B}. At high luminosity ($L_{\rm X}\gtrsim L_{\rm crit}$), the emitted radiation primarily escapes through the walls of the accretion column, forming a so-called ``fan beam'' \citep[][]{1973NPhS..246....1D}. At very low luminosity ($L_{\rm X}\ll L_{\rm crit}$), the radiation is
produced by hot spots or mounds at the polar cap and escapes roughly along magnetic field lines \citep[][]{1991ApJ...367..575B,1993ApJ...418..874N}, resulting a ``pencil beam''. In the intermediate luminosity range ($L_{\rm X}\lesssim L_{\rm crit}$), however, the emission pattern may be a hybrid combination of these two types \citep[][]{2000ApJ...529..968B}.

In this study, we investigate the 2020 giant outburst of the transient accreting pulsar 1A 0535+262 discovered by the \emph{Ariel \uppercase\expandafter{\romannumeral5}} satellite in 1975 with a pulsation period of $\sim$104 s \citep[][]{1975Natur.256..628R}. This system contains an O9.7\uppercase\expandafter{\romannumeral3}e donor star and a strongly magnetized NS \citep[][]{1998MNRAS.297L...5S,2023A&A...677A.134N}. The eccentricity and period of the binary orbit are $\sim$0.47 and $\sim$110.3 days, respectively \citep[][]{1996ApJ...459..288F}. The distance to this source is $\sim$2 kpc, which has been recently confirmed by \emph{Gaia} \citep[][]{2018AJ....156...58B}. Absorption line-like features at $\sim$50 and $\sim$100 keV were firstly detected with MirTTM/HEXE \citep[][]{1994A&A...291L..31K}, and have been confirmed by different missions and studies \citep[see e.g.][]{2005ATel..601....1K,2005ATel..605....1W,2005ATel..613....1I,2017A&A...608A.105B}. Several studies have investigated the $E_{\rm cyc}$-$L_{\rm X}$ relation with intriguing results. \citet[][]{2007A&A...465L..21C} reported that there were no significant changes of $E_{\rm cyc}$ with luminosity. On the contrary a possible positive $E_{\rm cyc}$-$L_{\rm X}$ correlation in a higher luminosity range was reported by \citet[][]{2015ApJ...806..193S}. Based on the pulse-to-pulse technique, \citet[][]{2011A&A...532A.126K} and \citet[][]{2013A&A...552A..81M} revealed a positive $E_{\rm cyc}$-$L_{\rm X}$ correlation on the short time scale of the pulse period. Recently, \citet[][]{2021ApJ...917L..38K} revealed both positive and negative $E_{\rm cyc}$-$L_{\rm X}$ correlations during the 2020 giant outburst based on the pulse-averaged spectroscopy derived from high-cadence observations of Insight-HXMT. Here, we focus on the luminosity dependence of cyclotron line energy during the 2020 giant outburst using pulse-to-pulse analysis in order to study luminosity changes on timescales of the pulse period and therefore extending/refining the range of luminosity. We stress that this is significantly different approach than that used in \citet[][]{2021ApJ...917L..38K}, and in fact new results are found.  We introduce the Insight-HXMT observations and data reduction in Section~\ref{sec:2}, present the data analysis and results in Section~\ref{sec:3}, discuss these in Section~\ref{sec:4} and finally summarize in Section~\ref{sec:5}.

\section{Observations and Data Reduction}\label{sec:2}
In the present study, we analyze the data from the Hard X-ray Modulation Telescope \citep[Insight-HXMT,][]{2014SPIE.9144E..21Z,2020SCPMA..6349502Z}, the first Chinese X-ray satellite launched on 2017 June 15 with a science payload that allows observations in a broad energy band (1--250 keV). Insight-HXMT include three collimated telescopes: the High Energy X-ray telescope (HE, NaI/CsI, 20--250 keV), the Medium Energy X-ray telescope (ME, Si pin detector, 5--30 keV), and the Low Energy X-ray telescope (LE, SCD detector, 0.7--13 keV), working in scanning and pointing observational modes and Gamma Ray Burst (GRB) mode. For details about the Insight-HXMT mission, see \citet[][]{2019ApJ...879...61Z,2020SCPMA..6349503L,2020SCPMA..6349504C,2020SCPMA..6349505C}. 

\begin{figure}
    \includegraphics[width=\columnwidth]{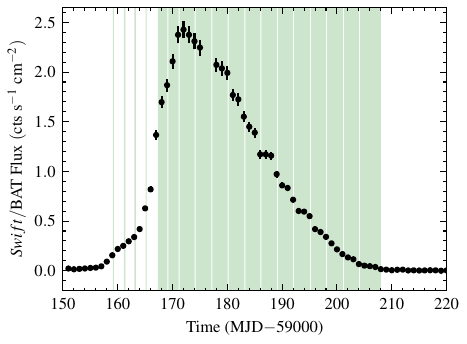}
    \caption{Swift/BAT light curve of 1A 0535+262 during the 2020 giant outburst with time time resolution of one point per day. The shaded areas indicate the time intervals of Insight-HXMT observations.}
    \label{fig:1}
\end{figure}

Fig.~\ref{fig:1} presents the light curve of 1A 0535+262 during its 2020 giant outburst from the Swift/BAT transient monitor \citep[][]{2013ApJS..209...14K}. 
The shaded areas represent the time intervals of the Insight-HXMT observations which well cover almost the entire outburst period. 
The Insight-HXMT Data Analysis Software v2.05, together with the current calibration model v2.06\footnote{\url{http://hxmtweb.ihep.ac.cn/software.jhtml}} and the standard Insight-HXMT Data Reduction Guide v2.05\footnote{\url{http://hxmtweb.ihep.ac.cn/SoftDoc.jhtml}}, are adopted in the extraction of the event data, under a series of criteria as recommended by the Insight-HXMT team: (1) the elevation angle (ELV) is larger than $10^{\circ}$; (2) the geometric cutoff rigidity (COR) is larger than 8 GeV; (3) the offset for the pointing position is less than $0.04^{\circ}$; (4) and the good time internals (GTIs) are at least 300 s away from the South Atlantic Anomaly (SAA).
In addition, we exclude the data of detector boxes affected by the Crab \citep[see also][]{2021ApJ...917L..38K,2022ApJ...932..106K,2022ApJ...935..125W}, since the Crab is located close to 1A 0535+262. 
The backgrounds are produced from blind detectors, using the tools: LEBKGMAP, MEBKGMAP and HEBKGMAP, version 2.0.9 based on the standard \emph{Insight-HXMT} background models \citep[][]{2020JHEAp..27...14L,2020JHEAp..27...24L,2020JHEAp..27...44G}. The XSPEC v12.12.0 software package \citep[][]{1996ASPC..101...17A} is used to perform the spectral analysis.

\section{Analysis and results}\label{sec:3}
\subsection{Pulse-to-pulse Analysis}
We produce pulse-amplitude-resolved spectra by applying the pulse-to-pulse technique elaborated in \citet[][]{2011A&A...532A.126K} and \citet[][]{2013A&A...552A..81M}.
This approach has recently employed to investigate the luminosity dependence of the CRSF energy in Cepheus X--4 and V 0332+53 using NuSTAR data \citep[see][]{2017A&A...601A.126V,2018A&A...610A..88V}. 
The technique enables the analysis of the spectral variability on the time scale of a pulse period. Furthermore, by segmenting a single time-averaged spectrum into multiple spectra at different luminosity levels, this method expands the luminosity range and increases the number of data points, providing detailed information regarding the luminosity dependence of the CRSF energy.

\begin{figure}
    \begin{minipage}[c]{0.46\textwidth}
\centering
    \includegraphics[width=\linewidth]{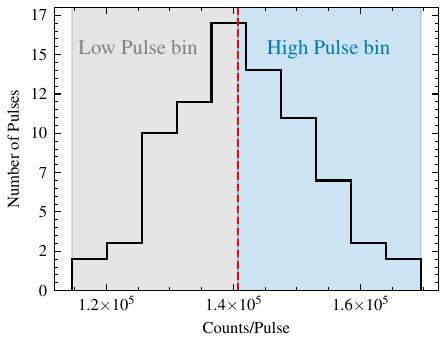}
\end{minipage}\\
\begin{minipage}[c]{0.46\textwidth}
\centering
    \includegraphics[width=\linewidth]{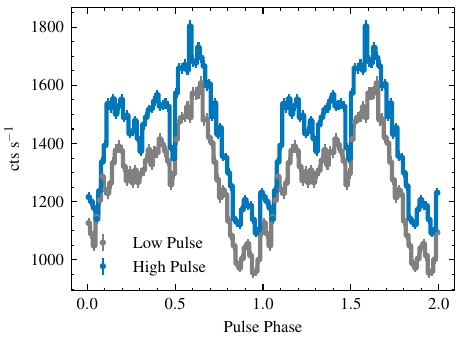}
\end{minipage}
    \caption{\textit{Top}: distributions of counts from ME in individual pulse (i.e. of pulse amplitudes) for the observation on MJD 59183, the red dashed line indicates the boundary of the amplitude bins, and colored shaded areas represent the counting ranges of low and high amplitude bins, respectively. \textit{Bottom}: averaged ME pulse profiles of higher (blue) and lower (gray) pulse-amplitude bins whose counting ranges are presented in the top panel.}
    \label{fig:AA}
\end{figure}

In our pulse-to-pulse analysis, we extract the net light curve from LE (1--10 keV), ME (10--30 keV) and HE (30--120 keV), respectively. The arrival times of photons are
corrected to the solar system barycentre with the HXMTDAS tool \textit{hxbary}, and the effects of binary orbital modulation are corrected using the ephemerides provided by \citet[][]{2012ApJ...754...20C}. Based on the binary- and barycentre-corrected data, we measure the pulse periods and produce pulse profiles \citep[see][for details about the timing analysis]{2022ApJ...935..125W, Hou2023}. Next, we measure the number of counts from ME during each pulse period, as its energy range occupies an intermediate position and typically encompasses the peak flux in the spectrum of an X-ray Pulsar (XRP). The number of counts summed over a pulse represents the pulse amplitude and is thus used as a measure of the pulse brightness. According to the measurement of the pulse amplitude in each pulse, we can build a frequency distribution of pulses as a function of the pulse amplitude (Fig.~\ref{fig:AA}). Based on the distribution of the pulse amplitude, the pulses can be divided into two bins at different brightness levels. As shown in the bottom panel of Fig.~\ref{fig:AA}, the pulse profiles in different amplitude bins indeed have significantly different count rates. The selection of the number of the bins mainly depends on the available counting statistics of the energy spectra, which should ensure that the cyclotron line energy can be well constrained in the spectral fitting. We stack the GTIs of pulses belonging to the same pulse-amplitude bin and subsequently extract the energy spectra from LE, ME, and HE, respectively, using the stacked GTIs. We defined a single observation as the combination of exposures within an one-day period. However, low-luminosity spectra exhibit poor counting statistics. To bolster the robustness of spectral fittings during low-luminosity intervals, we moderately relaxed the criteria for combining exposures during these stages. Specifically, for the outburst rising phase, exposures from MJD 59,163 to 59,165 are amalgamated into a single observation, while for the fading phase, exposures from MJD 59,204 to 59,207 are likewise merged into one observation. Following this, the pulse-to-pulse analysis is executed for each individual observation. We note that moderate changes in the combination strategy have small impact on the spectral outcomes. We have tried to build histograms of the pulse amplitude with a longer-time period, and the obtained spectral results are similar to the those from the original 1-day combination. However, constructing the amplitude histograms solely based on luminosity may introduce bias, since a time drift of the cyclotron line energy has been reported \citep[see][]{2021ApJ...917L..38K}. This suggests that there could be intrinsic differences in the spectra between the rising and fading phases.

\begin{figure*}
\centering
\begin{minipage}[c]{0.45\textwidth}
\centering
    \includegraphics[width=\linewidth]{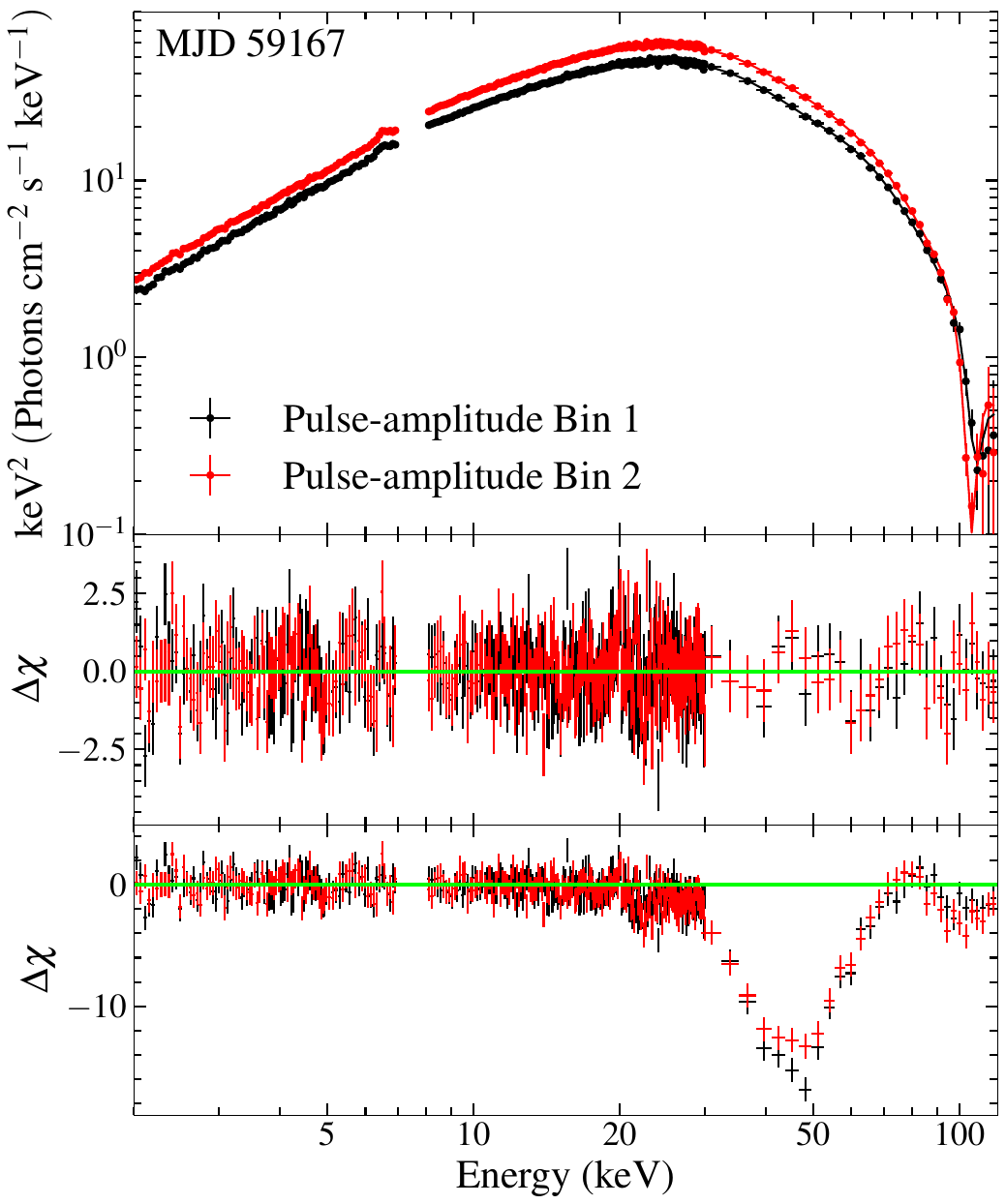}
\end{minipage}%
\begin{minipage}[c]{0.45\textwidth}
\centering
    \includegraphics[width=\linewidth]{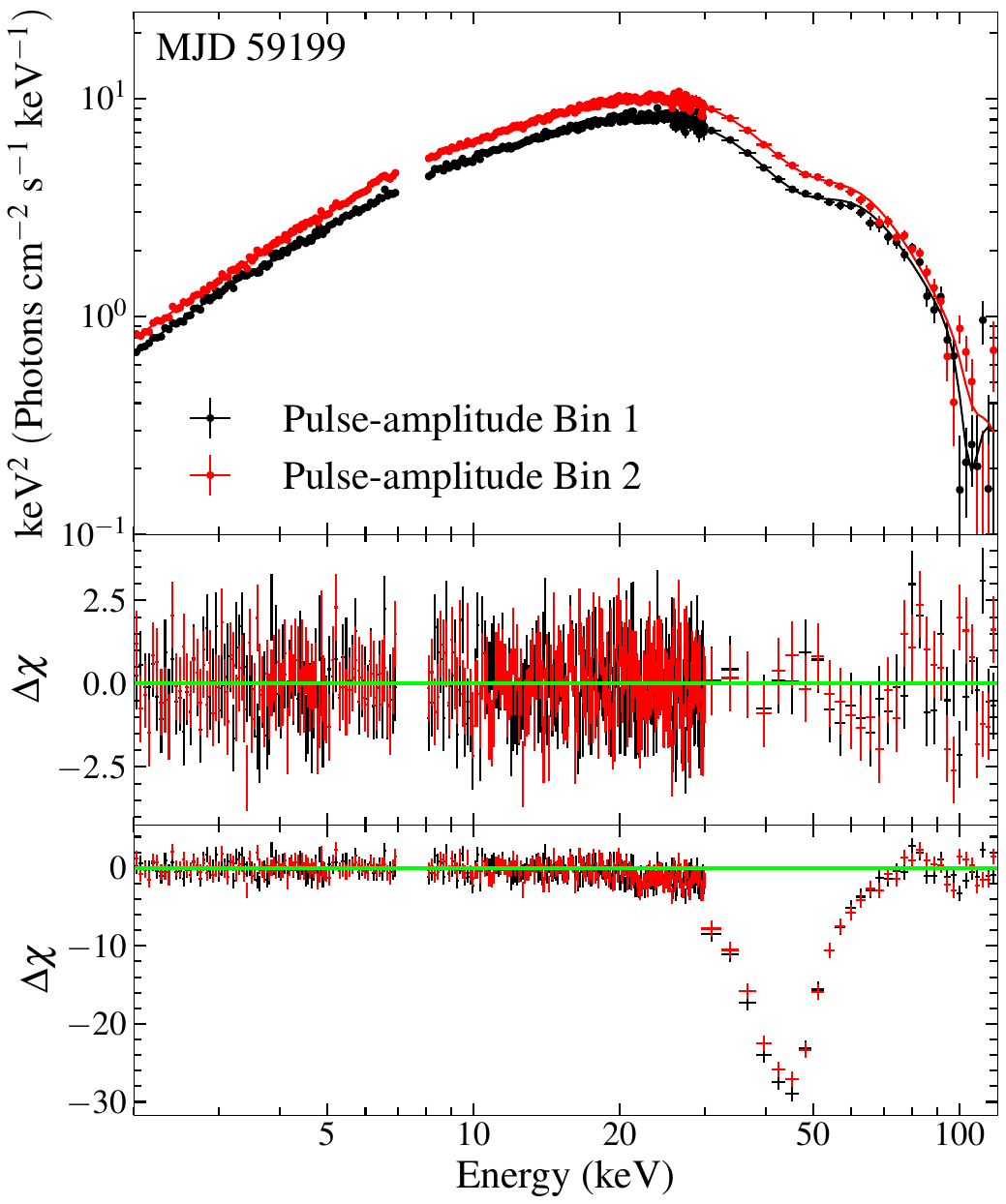}
\end{minipage}
    \caption{Unfolded pulse-amplitude energy spectra (the top panels) and the corresponding fitting residuals (the two bottom narrow panels) of MJD 59167 and MJD 59199. The spectral analysis is performed in the energy bands of 2--7 keV, 8--30 keV and 30--120 keV for LE, ME and HE, respectively. The middle panels show the best-fitting residuals with model: \textit{cons}$\times$\textit{tbabs}$\times$ \textit{mgabs}$\times$(\textit{bbodyrad}+\textit{bbodyrad}+\textit{Gaussian}+\textit{cutoffpl}), while the bottom panels display the residuals without the CRSF line model (\textit{mgabs}).}
    \label{fig:2}
\end{figure*}

In the spectral analysis, the energy bands adopted for the spectral fittings are 2--7 keV for LE, 8--30 keV for ME and 30--120 keV for HE. Considering the current accuracy of the instrument calibration, we add 0.5\%, 0.5\%, and 1\% systematic errors to the energy spectra for LE, ME, and HE, respectively. Following the recommendation of the Insight-HXMT calibration group, the spectra are rebinned as follows: (1) LE: channels 0--579 and 580--1535 are respectively grouped with 5 and 10 bins in each group; (2) ME: channels 0--1023 are grouped with 2 bins in each group; (3) HE: channels 0-255 are grouped with 2 bins in each group. With this grouping strategy, the spectra have at least 100 counts per energy bin, resulting in spectral data that closely resemble the Gaussian distribution. Additionally, the background spectra of the three instruments of Insight-HXMT are dependent on background models, which are mainly based on a large amount of the blank sky observations, so the spectral data of the background  are not Poissonian but Gaussian. As a result, $\chi^2$ statistics is used in the spectral analysis. In the present study, we use the same spectral model: \textit{constant}$\times$\textit{tbabs}$\times$ \textit{mgabs}$\times$(\textit{bbodyrad}+\textit{bbodyrad}+\textit{Gaussian}+\textit{cutoffpl}), as that used in the time-averaged spectral analysis \citep[][]{2021ApJ...917L..38K}, to fit each energy spectrum. In the fitting model, inter-calibration constants are required among the instruments. The constant value of the LE is fixed at 1, while the constant values of the ME and HE are allowed to vary. \textit{mgabs} is a multiplicative absorption model with a Gaussion profile,
\begin{align}
F'(E)&=F(E)\times mgabs=F(E)\notag\\
&\times\left[1-\tau_1 e^{-\frac{(E-E_{\rm cyc})^2}{2\sigma^2_1}}\right]\times\left[1-\tau_2 e^{-\frac{(E-R\times E_{\rm cyc})^2}{2\sigma^2_2}}\right],    
\end{align}
where $F'(E)$ is the spectrum modified by \textit{mgabs}, $E$ is the photon energy, $E_{\rm cyc}$ is the cyclotron line centroid energy, $\tau_1$ and $\sigma_1$ are the optical depth and the width of the line, respectively, $R$ is the centroid energy ratio of the first harmonic line and fundamental line, and $\tau_2$ and $\sigma_2$ are the optical depth and the width of the first harmonic line, respectively. To study the properties of the fundamental CRSF line energy, we fix the hydrogen column density ($N_{\rm H}$), the ratio of the first harmonic line and fundamental line of CRSFs ($R$), the width of the fundamental line of CRSFs ($\sigma_1$), the width of the first harmonic line of CRSFs ($\sigma_2$) and the width of the iron emission line ($\sigma_{\rm Fe}$) to $0.59\times10^{22}\ {\rm cm^{-2}}$, 2.3, 10 keV, 5 keV and 0.3 keV, respectively \citep[see also][]{2021ApJ...917L..38K}. We find each spectral fitting with the model obtains a satisfactorily reduced $\chi^2$. To compute the uncertainties at the 68\% confidence level,we employ the Markov Chain Monte Carlo (MCMC) method using the Goodman–Weare algorithm with 8 walkers and a total length of 40,000 \citep{2010CAMCS...5...65G}, and the initial 2000 elements are discarded as a burn-in period. We find that the autocorrelation length is typically 1000 elements, so the net number of independent samples in the parameter space we have is of the order of $10^4$. In order to further test the convergence of the MCMC chain, we compare the one- and two-dimensional projections of the posterior distributions for fitting parameters from the first and second halves of the chain, and we find no significant differences (see Fig.~\ref{fig:mcmc}). In Fig.~\ref{fig:2}, we display the representative pulse-amplitude-resolved spectra and present the corresponding best-fit spectral parameters of each model in Table~\ref{tab:1}. The best-fitting parameters of all spectra in this study are provided in Table~\ref{tab:A1}.

\begin{table*}
\caption{Best-fit Parameters of the representative Pulse-amplitude-resolved Spectra. The spectral analysis is performed in the energy bands of 2--7 keV, 8--30 keV and 30--120 keV for LE, ME and HE, respectively. The 0.5\%, 0.5\%, and 1\% systematic errors are added to the energy spectra for LE, ME, and HE, respectively. Uncertainties are reported at the 68\% confident level (1$\sigma$) using the MCMC technique, with length of 40,000.\label{tab:1}}
\begin{tabular}{cccccc}
\hline
Date &  & \multicolumn{2}{c}{2020-11-14} &
\multicolumn{2}{c}{2020-12-16} \\
\hline
MJD &  &
\multicolumn{2}{c}{59167} &
\multicolumn{2}{c}{59199} \\
\hline
Model & Parameters &
Pulse-amplitude Bin 1 & Pulse-amplitude Bin 2 &
Pulse-amplitude Bin 1 &
Pulse-amplitude Bin 2\\
\textit{tbabs} &
$N_{\rm H}$ ($10^{22}\ {\rm cm^{-2}}$) &
\multicolumn{4}{c}{0.59 (fixed)}\\
\textit{mgabs} &
$\sigma_1$ (keV) &
\multicolumn{4}{c}{10 (fixed)}\\
 & $\sigma_2$ (keV) &
\multicolumn{4}{c}{5 (fixed)}\\
 & $R$ &
\multicolumn{4}{c}{2.3 (fixed)}\\
\textit{Gaussian} &
$\sigma_{\rm Fe}$ (keV) &
\multicolumn{4}{c}{0.3 (fixed)}\\
\hline
\textit{mgabs} &
$E_{\rm cyc}$ (keV) &
$46.93^{+0.35}_{-0.25}$ & 
$45.99^{+0.36}_{-0.37}$ & 
$45.77^{+0.42}_{-0.37}$ &
$46.20^{+0.35}_{-0.31}$\\[5pt]
 & $\tau_1$ &
$0.17^{+0.01}_{-0.01}$ & 
$0.14^{+0.01}_{-0.01}$ & 
$0.38^{+0.02}_{-0.02}$ &
$0.34^{+0.01}_{-0.02}$\\[5pt]
 & $\tau_2$ &
$0.76^{+0.03}_{-0.22}$ & 
$0.93^{+0.13}_{-0.17}$ & 
$0.65^{+0.29}_{-0.23}$ &
$0.29^{+0.35}_{-0.15}$\\[5pt]
\textit{bbodyrad1} &
$kT$ (keV) &
$1.66^{+0.03}_{-0.03}$ & 
$1.73^{+0.04}_{-0.03}$ & 
$1.58^{+0.04}_{-0.05}$ &
$1.54^{+0.02}_{-0.03}$\\[5pt]
& Norm &
$84^{+7}_{-5}$ & 
$97^{+6}_{-6}$ &
$20^{+3}_{-2}$ &
$27^{+3}_{-3}$\\[5pt]
\textit{bbodyrad2} &
$kT$ (keV) &
$0.50^{+0.02}_{-0.01}$ & 
$0.55^{+0.01}_{-0.01}$ & 
$0.47^{+0.02}_{-0.03}$ &
$0.37^{+0.02}_{-0.02}$\\[5pt]
& Norm &
$2945^{+283}_{-380}$ & 
$2781^{+251}_{-201}$ & 
$492^{+132}_{-122}$ &
$2287^{+982}_{-617}$\\[5pt]
\textit{Gaussian} &
$E_{\rm Fe}$ (keV) &
$6.56^{+0.03}_{-0.03}$ & 
$6.56^{+0.04}_{-0.02}$ & 
$6.57^{+0.12}_{-0.13}$ &
$6.45^{+0.07}_{-0.10}$\\[5pt]
& Norm ($10^{-2}$) &
$2.8^{+0.3}_{-0.3}$ & 
$3.6^{+0.2}_{-0.3}$ & 
$0.3^{+0.1}_{-0.1}$ &
$0.6^{+0.1}_{-0.1}$\\[5pt]
\textit{cutoffpl} &
$\Gamma$ &
$-0.18^{+0.01}_{-0.02}$ & $-0.29^{+0.01}_{-0.01}$ & 
$0.37^{+0.02}_{-0.02}$ &
$0.30^{+0.02}_{-0.02}$\\[5pt]
& $E_{\rm fold}$ (keV) &
$11.79^{+0.04}_{-0.09}$ & $11.30^{+0.07}_{-0.05}$ & 
$15.86^{+0.20}_{-0.18}$ &
$15.03^{+0.13}_{-0.20}$\\[5pt]
& Norm &
$0.32^{+0.01}_{-0.01}$ & $0.30^{+0.01}_{-0.01}$ & 
$0.20^{+0.01}_{-0.01}$ &
$0.21^{+0.01}_{-0.01}$\\[5pt]
Luminosity &
$L_{2-150{\rm keV}}\ (10^{37}\ {\rm erg\ s^{-1}})$ & 
$6.60\pm0.02$ & 
$8.13\pm0.03$ & 
$1.28\pm0.01$ &
$1.53\pm0.01$\\
\hline
Fitting & $\chi^2$/d.o.f &
288.86/298 & 
263.67/298 &
287.59/298 &
283.89/298\\
\hline
\end{tabular}
\end{table*}

\subsection{Luminosity Dependence of the CRSF Centroid Energy}
\begin{figure*}
\centering
\begin{minipage}[c]{0.48\textwidth}
\centering
    \includegraphics[width=\linewidth]{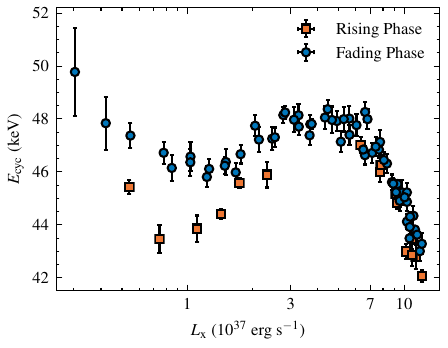}
\end{minipage}%
\begin{minipage}[c]{0.48\textwidth}
\centering
    \includegraphics[width=\linewidth]{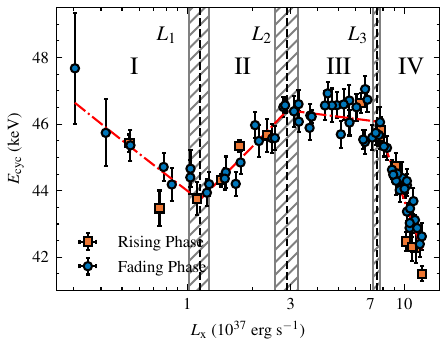}
\end{minipage}
    \caption{\textit{Left}: Luminosity ($L_{\rm X}$) dependence of the cyclotron line centroid energy ($E_{\rm cyc}$) derived from the pulse-to-pulse analysis, where the uncertainties are reported at the 68\% confident level (1$\sigma$). The data points from the rising and fading phases are plotted as orange squares and blue circles, respectively. \textit{Right}: Luminosity dependence of cyclotron line centroid energy with the linear energy drift of $\dot{E}_{\rm cyc}=0.047\ {\rm keV\ day^{-1}}$ taken into account (symbols are the same as in the left panel). The red dashed line represents the best fitting result with the broken power-law model which has three break luminosities ($L_1$, $L_2$ and $L_3$). These three vertical dashed lines and shaded ares are the best fitting break luminosities and corresponding uncertainties, respectively. These transitional luminosities suggest the division of the $E_{\rm cyc}$-$L_{\rm X}$ relation into four zones (Zones \uppercase\expandafter{\romannumeral1}, \uppercase\expandafter{\romannumeral2}, \uppercase\expandafter{\romannumeral3}, and \uppercase\expandafter{\romannumeral4}) with different types of the $E_{\rm cyc}$-$L_{\rm X}$ correlation.}
    \label{fig:3}
\end{figure*}

Based on spectral fittings of the pulse-amplitude-resolved spectra, we calculate the X-ray luminosity ($L_{\rm X}$) in the 2--150 keV energy range, assuming the source is at a distance of 2 kpc. To achieve this, we extend the energy response of the model from 0.1 to 200 keV using the $energies$ command in XSPEC. Subsequently, the model fluxes in the 2--150 keV energy range and their corresponding uncertainties are computed for the three instruments using the $flux$ command in XSPEC when the MCMC chains are loaded. This allows for the derivation of the flux distribution from the MCMC chains. The luminosities reported here represent average values derived from the three instruments, and the associated errors are calculated using standard error propagation formula. It is worth noting that the reported luminosity is not corrected for absorption. However, this does not affect the luminosity dependence of $E_{\rm cyc}$, since the value of $N_{\rm H}$ is fixed in the fitting. Additionally, we find that the systematic difference between the values of unabsorbed flux and that linked to absorption is less than 1\%, primarily due to the relatively small distance of the source. The left panel of Fig.~\ref{fig:3} displays the luminosity dependence of the CRSF centroid energy, with orange squares and blue circles representing the rising and fading phases of the outburst, respectively.
As a result of the pulse-to-pulse analysis, the number of data points shown in the left panel of Fig.~\ref{fig:3} is twice as large as the result from the pulse-averaged spectral analysis \citep[see][]{2021ApJ...917L..38K}, enabling us to study the $E_{\rm cyc}$-$L_{\rm X}$ relations in greater detail.
Our analysis confirms the previously observed anti-correlation between $E_{\rm cyc}$ and $L_{\rm X}$ at high luminosity when ($L_{\rm X}\gtrsim7\times10^{37}\ {\rm erg\ s^{-1}}$). However, at lower luminosity, we find a more complex relationship between $E_{\rm cyc}$ and $L_{\rm X}$. Specifically, in the luminosity range of $\sim(3$--$7)\times10^{37}\ {\rm erg\ s^{-1}}$, $E_{\rm cyc}$ is not correlated with $L_{\rm X}$. In the narrow luminosity range of $\sim(1$--$3)\times10^{37}\ {\rm erg\ s^{-1}}$, $E_{\rm cyc}$ is positively correlated with $L_{\rm X}$ during both the rising and fading phases. At the lowest luminosity level ($L_{\rm X}\lesssim1\times10^{37}\ {\rm erg\ s^{-1}}$), the positive $E_{\rm cyc}$-$L_{\rm X}$ correlation appears to break, and $E_{\rm cyc}$ shows an increasing trend with decreasing $L_{\rm X}$ during both the rising and fading phases. Such a complex behaviour is reported here for the first time. 

It is also important to note that the measured $E_{\rm cyc}$ during the rising phase is systematically lower than that during the fading phase, a behavior that has been reported in the pulse-averaged spectral analysis by \citet[][]{2021ApJ...917L..38K}. Similarly, line energy differences between the two outburst stages were detected in V 0332+53 during its 2015 giant outburst \citep[][]{2016MNRAS.460L..99C,2017MNRAS.466.2143D}. However, in the case of 1A 0535+262, contrary to what reported in literature, $E_{\rm cyc}$ during the rising phase is smaller than during the fading phase. Nevertheless, the trends in evolution between the two phases are nearly parallel. 

To study changes in the $E_{\rm cyc}$-$L_{\rm X}$ relation by combining data points from the rising and fading phases, we follow the method of \citet[][]{2017MNRAS.466.2143D}, \citet[][]{2018A&A...610A..88V}, and \citet[][]{2021ApJ...917L..38K}, assuming that $E_{\rm cyc}$ undergoes a linear time drift ($\dot{E}_{\rm cyc}$) during the outburst. Specifically, we assume that each observation probes an ``instantanous'' changes in the cyclotron line energy. However, the influence of the time variation is negligible within individual observations. Then, for each data point present in the left panel of Fig. 4, $E_{\rm cyc}$ is adjusted by subtracting a factor of $\dot{E}_{\rm cyc}\times \Delta t$, where $\Delta t$ represents the difference between the reference time (MJD 59,159) and the time of observation. We then use a broken power-law model with three break luminosities to fit the data points with the linear energy drift of $\dot{E}_{\rm cyc}$ taken into account. This model has nine free parameters: four power-law indices ($\Gamma_1$, $\Gamma_2$, $\Gamma_3$, and $\Gamma_4$), three break luminosities ($L_1$, $L_2$, and $L_3$), normalization ($K$), and the linear time drift of line energy, $\dot{E}_{\rm cyc}$. The model fitting is performed by using the MCMC method with the Goodman-Weare algorithm with 8 walkers and a total length of 40,000. The uncertainty in $E_{\rm cyc}$ at each data point is accounted for  using a heteroscedastic Gaussian likelihood, and the uncertainty in $L_{\rm X}$ is accounted for by incorporating a sampling process during the execution of the MCMC chain.
The luminosity dependence of $E_{\rm cyc}$, corrected for $\dot{E}_{\rm cyc}$, is presented in the right panel of Fig.~\ref{fig:3}, with the best-fitting result using the broken power-law model plotted as the red dashed line. The model yields luminosity break values of $L_1=(1.14\pm0.12)\times10^{37}\ {\rm erg\ s^{-1}}$, $L_2=(2.89\pm0.35)\times10^{37}\ {\rm erg\ s^{-1}}$, and $L_3=(7.47\pm0.26)\times10^{37}\ {\rm erg\ s^{-1}}$. These transitional luminosity levels suggest the division of the $E_{\rm cyc}$-$L_{\rm X}$ relation into four zones (Zones \uppercase\expandafter{\romannumeral1}, \uppercase\expandafter{\romannumeral2}, \uppercase\expandafter{\romannumeral3}, and \uppercase\expandafter{\romannumeral4}) with different types of the $E_{\rm cyc}$-$L_{\rm X}$ correlation. For the linear time drift, $\dot{E}_{\rm cyc}$, we obtain an increase rate of $0.046\pm0.004\ {\rm keV\ day^{-1}}$ for cyclotron line energy, consistent with the result of \citet[][]{2021ApJ...917L..38K}.
We also find that the best-fitting statistics of the broken power-law model with only one break at $L_3$ improve from $\chi^2=219.62$ for 73 degrees of freedom (d.o.f.) to $\chi^2=117.7$ for 69 d.o.f. for the presented three-breaks broken power-law model. This indicates that the data preferred the model with three breaks at $\sim5.8\sigma$ confidence based on the F-test.
Performing a similar comparison between broken power-law models with two breaks at $L_1$ and $L_3$ and three breaks, the F-test gives $\sim4.9\sigma$ confidence for the three breaks. This implies that the data prefer that the $E_{\rm cyc}$-$L_{\rm X}$ relation changes at $L_2$ at $\sim4.9\sigma$ confidence. It is also important to note that the estimated confidence level can be affected by the choice of fitting model and the assumption that the CRSF centroid energy exhibits a linear drift during the outburst.
\begin{figure*}
\centering
\begin{minipage}[c]{0.48\textwidth}
\centering
    \includegraphics[width=\linewidth]{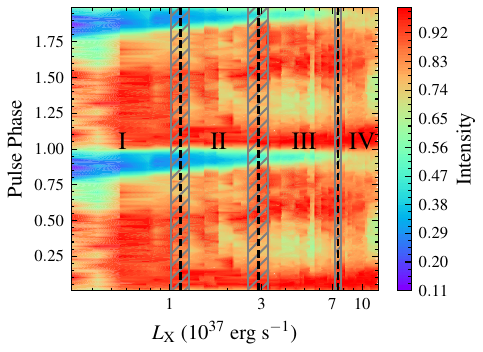}
\end{minipage}%
\begin{minipage}[c]{0.48\textwidth}
\centering
    \includegraphics[width=\linewidth]{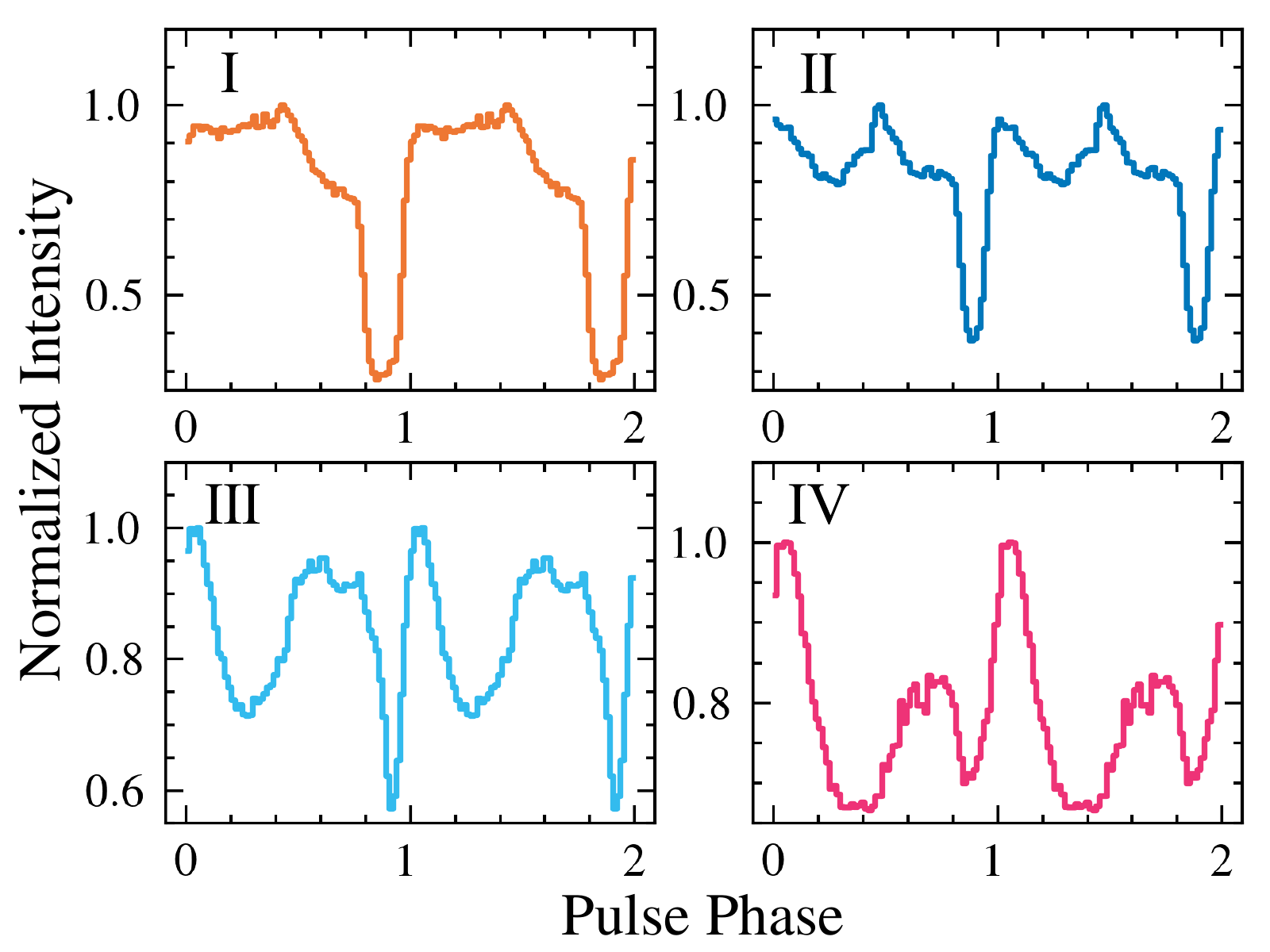}
\end{minipage}
    \caption{\textit{Left}: Luminosity dependence of pulse profiles in the energy band of 30--120 keV during the whole outburst. The color bar displays the intensity of the pulse profile normalized in the 0--1 range. These three vertical dashed lines are the break luminosities ($L_1$, $L_2$ and $L_3$) shown in the left panel of Fig.~\ref{fig:3}, and the shaded areas are the corresponding uncertainties at 1$\sigma$ confidence level. These transitional luminosities suggest the division of the $E_{\rm cyc}$-$L_{\rm X}$ relation into four zones (Zones \uppercase\expandafter{\romannumeral1}, \uppercase\expandafter{\romannumeral2}, \uppercase\expandafter{\romannumeral3}, and \uppercase\expandafter{\romannumeral4}). \textit{Right}: Example pulse-amplitude-averaged profiles for the four distinct zones. These example pulse profiles for the four zones are from observations of MJD 59202, 59196, 59188 and 59173, respectively.}
    \label{fig:4}
\end{figure*}

\subsection{Relationship Between the CRSF energy and Pulse Profiles}
The Insight-HXMT data have been utilized for providing detailed timing analyses of 1A 0535+262 during its 2020 giant outburst, as reported by \citet[][]{2022A&A...659A.178R,2022ApJ...935..125W,2022MNRAS.517.1988M}. Theoretical models suggest that the beam pattern changes across different accretion regimes \citep[e.g.][]{1976MNRAS.175..395B,2012A&A...544A.123B}. Therefore, analyzing the correlation between $E_{\rm cyc}$ and $L_{\rm X}$, along with the pulse profile, can offer valuable insights into state transitions. In general, high-energy photons mainly originate from the polar caps through direct emission, while low-energy photons include contributions from direct emission, thermal components, and reflection on the NS surface \citep[see e.g.][]{2023ApJ...945..138H,2023MNRAS.526.3637L}. Consequently, the pulse profile in the high-energy range is more effective in tracing beam pattern transitions. Here, we average the pulse profiles in the high-energy bands of 30--120 keV within each pulse bin according to the pulse-to-pulse analysis performed in Section~\ref{sec:2}, resulting in the amplitude-resolved pulse profiles (referred to as pulse profiles hereafter). The left panel of Fig.~\ref{fig:4} displays the luminosity dependence of the pulse profile, with the color bar representing the normalized intensity. Based on the evolution of the pulse-averaged pulse profiles, \citet[][]{2022ApJ...935..125W} suggested a transition luminosity of nearly $1.1\times10^{37}\ {\rm erg\ s^{-1}}$. We note that our results from the pulse-to-pulse analysis are consistent with those of \citet[][]{2022ApJ...935..125W}, as we find a transition luminosity close to this value (see Figure~\ref{fig:4}). Interestingly, such a transition luminosity aligns with $L_1$, which represents the lowest break luminosity in the $E_{\rm cyc}$-$L_{\rm X}$ correlation. By examining the luminosity dependence of the pulse profile and that of $E_{\rm cyc}$ together, we find a co-evolution between these two aspects. At luminosities below $\sim L_1=(1.14\pm0.12)\times10^{37}\ {\rm erg\ s^{-1}}$, $E_{\rm cyc}$ exhibits to be anti-correlated with $L_{\rm X}$, and the pulse profile in the high energy bands (30--120 keV) appears as a single broad peak. In the luminosity range of $\sim(1-3)\times10^{37}\ {\rm erg\ s^{-1}}$ ($L_1\lesssim L_{\rm X}\lesssim L_2$), where the $E_{\rm cyc}$-$L_{\rm X}$ correlation is positive, the pulse profile changes from a single-peak structure to a double-peak one. When $L_2\lesssim L_{\rm X}\lesssim L_3$, the pulse profile exhibits a significant double-peak structure, and $E_{\rm cyc}$ remains relatively constant. In Zone \uppercase\expandafter{\romannumeral4}, the pulse profile still shows a double-peak structure, but the two peaks in one cycle have significantly different intensities, and $E_{\rm cyc}$ shows an anti-correlation with luminosity. To clearly illustrate the evolution of the pulse profile across different zones, we present example pulse-amplitude-averaged profiles for the four zones in the right panel of Fig.~\ref{fig:4}.

\section{Discussion}\label{sec:4}
We have studied the luminosity dependence of the cyclotron line energy of 1A 0535+262 during its 2020 giant outburst. Using high-cadence observations from Insight-HXMT, we performed the pulse-to-pulse analysis to reveal the comprehensive scenario of the $E_{\rm cyc}$-$L_{\rm X}$ relation in a broad luminosity range of $\sim(0.03$--$1.3)\times10^{38}\ {\rm erg\ s^{-1}}$. Specifically, through the use of a large number of data points obtained from the pulse-to-pulse analysis, we were able to fit the $E_{\rm cyc}$-$L_{\rm X}$ correlation with a broken power-law model. Based on the fitting, we identified three transitional luminosity levels ($L_1$, $L_2$, and $L_3$) with a confidence level of  $\sim5\sigma$. These transitional luminosity levels suggest the division of the $E_{\rm cyc}$-$L_{\rm X}$ relation into four zones (Zones \uppercase\expandafter{\romannumeral1}, \uppercase\expandafter{\romannumeral2}, \uppercase\expandafter{\romannumeral3}, and \uppercase\expandafter{\romannumeral4}) with different types of the $E_{\rm cyc}$-$L_{\rm X}$ correlation. Our results confirm the recently reported anti-correlation between $E_{\rm cyc}$ and $L_{\rm X}$ in the high luminosity range (Zone \uppercase\expandafter{\romannumeral4}, 
i.e. $L_{\rm X}\gtrsim L_3$) from pulse-averaged spectral analysis \citep[see][]{2021ApJ...917L..38K}. However, we found that the $E_{\rm cyc}$-$L_{\rm X}$ relation at the lower luminosities ($L_{\rm X}\lesssim L_3$) is more complex, and the pulse profile also shows significant changes. In Zone \uppercase\expandafter{\romannumeral3} ($L_2\lesssim L_{\rm X}\lesssim L_3$), $E_{\rm cyc}$ remains roughly constant, and the pulse profile in the high energy bands (30--120 keV) displays a double-peak structure. In Zone \uppercase\expandafter{\romannumeral2} ($L_1\lesssim L_{\rm X}\lesssim L_2$), $E_{\rm cyc}$ shows positive correlation with $L_{\rm X}$, and the pulse profile exhibits a transition between double-peak and single-peak structure. At the lowest luminosity level (Zone \uppercase\expandafter{\romannumeral1}, i.e. $L_{\rm X}\lesssim L_1$), $E_{\rm cyc}$ seems to be anti-correlated with $L_{\rm X}$, and the pulse profile becomes a pure single broad peak.

The variation of $E_{\rm cyc}$ with pulse phase is generally believed to be due to the changing viewing angle under which the emission regions are seen \citep{2007A&A...465L..25S}. A phase-resolved analysis of this source with Insight-HXMT data have been conducted by \citet{2022ApJ...932..106K}. They showed that the transition of $E_{\rm cyc}$-$L_{\rm X}$ correlation at the high luminosity level ($\sim6\times10^{37}\ {\rm erg\ s^{-1}}$) identified in our analysis was also valid when certain pulse phases were analyzed. However, the remarkable phase dependency of $E_{\rm cyc}$ variations above $\sim6\times10^{37}\ {\rm erg\ s^{-1}}$ were observed \citep[see Fig. 3 of][]{2022ApJ...932..106K}, which means the phase variations of $E_{\rm cyc}$ might potentially affect the obtained $E_{\rm cyc}$-$L_{\rm X}$ correlation based on the phase-averaged spectral fitting in Zone \uppercase\expandafter{\romannumeral4}. Nevertheless, it appears that the influence of this phase dependency is relatively weak at luminosities in Zones \uppercase\expandafter{\romannumeral1}, \uppercase\expandafter{\romannumeral2} and \uppercase\expandafter{\romannumeral3}. To provide a more specific discussion, in this study, we disregard this effect for a pulse-to-pulse analysis with higher counting statistics and focus on studying the pulse-amplitude-resolved spectra averaged over the pulse profile.

\subsection{Cyclotron line behaviour at $L_{\rm X}>10^{37}\,{\rm erg\,s^{-1}}$}
Since the work of \citet[][]{1976MNRAS.175..395B}, it is generally accepted that the accretion mode of an accreting pulsars changes between sub- and supercritical regimes with the mass accretion rate, $\dot{M}$, hence with luminosity. 
The transition between the two accretion regimes occurs at the critical luminosity, $L_{\rm crit}$, which is determined by the properties of the neutron star \citep[e.g., magnetic field strength,][]{1976MNRAS.175..395B,2012A&A...544A.123B,2015MNRAS.447.1847M,2015MNRAS.454.2539M}. In the supercritical regime, where $L_{\rm X}\gtrsim L_{\rm crit}$, the radiation pressure dominates the gas deceleration, leading to the formation of an accretion column whose height could be positively correlated with the mass accretion rate. 
In this case, the energy of CRSF forming in accretion column \citep{2007AstL...33..368T} or at the stellar surface due to the reflection and reprocessing of X-ray photons \citep{2013ApJ...777..115P,2015MNRAS.448.2175L} is expected to be anti-correlated to luminosity. 

At $L_{\rm X}\lesssim L_{\rm crit}$, the accreting gas is decelerated in the atmosphere of a neutron star \citep{1969SvA....13..175Z} or at some height either via Coulomb interactions \citep[][]{2007A&A...465L..25S,2012A&A...544A.123B}, or through a collissionless shock above the stellar surface \citep[][]{1982ApJ...257..733L,2004AstL...30..309B}.
These scenarios require the positive correlation between the cyclotron line energy and accretion luminosity.
In the case of accretion flow deceleration in the atmosphere of a neutron star, the positive correlation is due to the Doppler effect in the accretion channel above the polar caps (see e.g., \citealt{2015MNRAS.454.2714M}).
In the case of Coulomb interactions or collisionless shock above the stellar surface, the positive correlation is due to the expected variations of emission region height above the surface (see Discussion in \citealt{2017MNRAS.466.2752R}).

The transition of X-ray pulsars between sub- and supercritical states is expected to be associated with the changes of a beam pattern forming close to the neutron star surface \citep{1973A&A....25..233G}.
Beam pattern variations change the typical angle between the field direction and the momentum of photons leaving the line-forming region.
It can affect the line centroid energy \citep{2014ApJ...781...30N} and possibly complicates the transition between the positive and negative correlation at $L_{\rm X}\sim L_{\rm crit}$.

As presented in Figs~\ref{fig:3} and~\ref{fig:4}, the cyclotron line energy is positively correlated to $L_{\rm X}$ in Zone \uppercase\expandafter{\romannumeral2}, where $L_{\rm X}\sim(1$--$3)\times10^{37}\ {\rm erg\ s^{-1}}$. Conversely, the sign of the correlation turns negative when $L_{\rm X}\gtrsim L_3$ (Zone \uppercase\expandafter{\romannumeral4}). These two luminosity ranges could potentially signify sub-critical and supercritical regimes, respectively. Before the 2020 outburst of 1A 0535+262, the occurrence of both positive and negative $E_{\rm cyc}$-$L_{\rm X}$ correlations in an outburst was a rare phenomenon, previously observed only in V 0332+53 \citep[][]{1990ApJ...365L..59M,2006MNRAS.371...19T,2016MNRAS.460L..99C,2017MNRAS.466.2143D,2018A&A...610A..88V}. It is worth noting that previous studies of 1A 0535+262 at luminosities below $L_3$ yielded inconclusive results; some claimed a positive correlation \citep{2015ApJ...806..193S,2021ApJ...917L..38K}, while others reported the absence of any correlation \citep{2006MNRAS.371...19T,2007A&A...465L..21C,2013ApJ...764L..23C,2013A&A...552A..81M}. The complexity of these findings may be attributed to the relatively narrow luminosity range of the positive correlation in this source, making it challenging to detect. Thanks to the high-cadence observational strategy and a pulse-to-pulse technique, the positive $E_{\rm cyc}$-$L_{\rm X}$ correlation can now be robustly observed by Insight-HXMT. It is important to note that the observed luminosity range of the positive $E_{\rm cyc}$-$L_{\rm X}$ correlation in this study is consistent with that reported by \citet{2015ApJ...806..193S} using \textit{INTEGRAL}/SPI observations. Moreover, we observe that $E_{\rm cyc}$ appears to exhibit a plateau in an intermediate luminosity range (Zone \uppercase\expandafter{\romannumeral3}) between Zones \uppercase\expandafter{\romannumeral2} and \uppercase\expandafter{\romannumeral4} as determined by fitting with the broken power-law model (see Fig.~\ref{fig:3}). 
This plateau of the $E_{\rm cyc}$-$L_{\rm X}$ correlation observed in the luminosity range of $L_2$--$L_3$ constitutes a novel observational phenomenon, potentially offering new perspectives for theoretical studies to explore the complex physical processes within the accretion column. A possible explanation for the plateau could be that when $L_1 \lesssim L_{\rm X} \lesssim L_2$, the source enter a mixed regime. In this mixed regime, it is possible that both Coulomb interactions and radiative pressure play crucial roles in decelerating the infalling gas within the accretion column. Consequently, the radiative height of the CRSF might exhibit relative stability within this luminosity range. While in Zone \uppercase\expandafter{\romannumeral2}, Coulomb interactions dominate the deceleration, resulting in the positive correlation between $E_{\rm cyc}$ and $L_{\rm X}$. In this scenario, the so-called critical luminosity might be located in the Zone \uppercase\expandafter{\romannumeral3}.

\subsection{Cyclotron line behaviour at $L_{\rm X}<10^{37}\,{\rm erg\,s^{-1}}$}
At the lowest luminosity level (Zone I, $L_{\rm X}\lesssim1\times10^{37}\ {\rm erg\ s^{-1}}$), the early researches reported  less changes of the cyclotron line energy \citep[e.g.][]{2006ApJ...648L.139T,2007A&A...465L..21C,2013ApJ...764L..23C}. Additionally, a detailed analysis of a deep NuSTAR observation at a low luminosity of $\sim7\times10^{34}\ {\rm erg\ s^{-1}}$ by \citet{2019MNRAS.487L..30T} excluded a positive correlation at low luminosities of this source. In this study, we observe a break in the positive $E_{\rm cyc}$-$L_{\rm X}$ correlation when $L_{\rm X}<L_1$. Furthermore, the pulse profile displays a single broad peak in this luminosity range (see Fig.~\ref{fig:4}). It is important to note that a similar transition of the correlation during the 2011 outburst of 1A 0535+262 have been reported by \citet{2015ApJ...806..193S} using \textit{INTEGRAL}/SPI observations, who claimed that the transition occurred between $6.8\times10^{36}$ and $1.2\times10^{37}\ {\rm erg\ s^{-1}}$, likely corresponding to the Coulomb-stopping limit luminosity \citep{2012A&A...544A.123B}. They also detected changes in the pulse profile during the transition \citep[see Fig. 12 of][]{2015ApJ...806..193S}. Our correlation fitting suggests a transition luminosity of $\sim1.14\times10^{37}\ {\rm erg\ s^{-1}}$, consistent with their result. In comparison with this previous work, we used more observations and a novel technique to robustly confirm such a transition at a $\sim5\sigma$ confidence level.

In the sub-critical regime, if the final deceleration of the in-falling gas occurs via Coulomb interactions, ones expect there could be a Coulomb-stopping limit luminosity \citep[see][for details]{2012A&A...544A.123B}, $L_{\rm coul}\approx1.17\times10^{37}\times (B_*/10^{12}\ {\rm G})^{-1/3}\ {\rm erg\ s^{-1}}$, where $B_*$ is the magnetic field strength at the stellar surface. 
For 1A 0535+262, the detected $E_{\rm cyc}$ is around 46 keV, so we can estimate $L_{\rm coul}$ at $\sim0.74\times10^{37}\ {\rm erg\ s^{-1}}$. 
We note that the estimated $L_{\rm coul}$ is close to the lowest transition luminosity, $L_1=(1.14\pm0.12)\times10^{37}\ {\rm erg\ s^{-1}}$. 
However, it is not very conclusive what happens when $L_{\rm X}\lesssim L_{\rm coul}$. 
A possible scenario is that the gas deceleration occurs via passage through a gas-mediated shock near the NS surface \citep[][]{1982ApJ...257..733L}, forming a small accretion mound \citep{2013MNRAS.430.1976M}. 
In this case, the emission region approaches the stellar surface quickly as $L_{\rm X}$ is reduced and the radiative beam pattern consists of pencil component only \citep[][]{1991ApJ...367..575B,1993ApJ...418..874N}. 
When $L_{\rm X}$ decreases, the change in the sign of the $E_{\rm cyc}$-$L_{\rm X}$ correlation from positive to negative has also been observed in GRO J1008-57 during its 2017 outburst \citep{2021ApJ...919...33C}. 
However, the pulse profile in GRO J1008-57 does not exhibit significant changes during the transition of the $E_{\rm cyc}$-$L_{\rm X}$ correlation \citep{2020ApJ...899L..19G}. 
Consequently, \citet{2021ApJ...919...33C} proposed variations in the accretion channel geometry to explain the observed anti-correlation between $E_{\rm cyc}$ and $L_{\rm X}$ at lower luminosity levels. 
This is attributed to the fact that a reduction in the length of the accretion channel could potentially lead to a smaller redshift of the line energy by increasing the local X-ray energy flux, even as the total luminosity decreases. 

Another explanation of the negative correlation at low luminosity level is related to the specific structure of NS atmosphere.
If the accretion flow is stopped by the Coulomb collisions in the upper atmospheric layers, one would expect an inverse temperature profile with the overheated optically layers and relatively cold underlying atmosphere \citep{1969SvA....13..175Z}.
At very low mass accretion rates, this atmospheric structure results in the appearance of two components of broadband energy spectra: the low-energy one related to the thermal emission of the atmosphere and the high-energy one due to emission and further reprocessing of cyclotron photons due to magnetic Compton scattering and absorption (see, e.g., \citealt{2021MNRAS.503.5193M,2021A&A...651A..12S}).
This spectral transition was reported in a few accreting pulsars, including 1A~0535+262 \citep{2019MNRAS.483L.144T,2019MNRAS.487L..30T}. However, in this study, there is no evidence of an additional high-energy component, as reported by \cite{2019MNRAS.487L..30T} at low luminosities below $10^{36}\ {\rm erg\ s^{-1}}$, appearing in spectra. We suggest this may because all Insight-HXMT observations on this source were conducted at luminosities beyond this level.
According to the numerical simulations based on the Monte Carlo approach, the cyclotron absorption feature tend to be shifted towards higher energies with respect to the cyclotron resonance energy in the atmosphere \citep{2021MNRAS.503.5193M}.
Thus, the negative correlation between the cyclotron line energy and luminosity observed at low mass accretion rates can be related to the changes in the atmospheric structure.

An alternative explanation could be that when the accretion rate is extremely low ($L_{\rm X}\lesssim10^{-4}L_{\rm crit}$), one would expect no gas-mediated shock above the NS and the optical thickness of the accretion channel becomes low even for cyclotron photons and the scattering feature is formed in the NS atmosphere (see Section 3.2 in \citealt{2015MNRAS.454.2714M}). 
In this situation, there would be no redshift of the line energy, and the line energy is expected to exhibit an anti-correlation with the luminosity \citep{2021ApJ...919...33C}. However, the transitional luminosity, $L_1$ ($\sim10^{37}\ {\rm erg\ s^{-1}}$), observed in 1A 0535+262 is at least three orders of magnitude higher than what is anticipated in this scenario. 

\subsection{Cyclotron line energy drift}
As already mentioned in Sec.~\ref{sec:3}, $E_{\rm cyc}$ in the rising phase is systematically lower than that in the fading phase, although the behaviors of the $E_{\rm cyc}$-$L_{\rm X}$ relation from the two stages are similar \citep[see also][]{2021ApJ...917L..38K}. The difference of $E_{\rm cyc}$ between the two outburst stages could be corrected by introducing a phenomenological increase rate of 0.046 keV d$^{-1}$ to the cyclotron line energy. Differences in $E_{\rm cyc}$ from the two stages of an outburst have been also found in the 2015 outburst of V 0332+53 \citep{2016MNRAS.460L..99C,2017MNRAS.466.2143D,2018A&A...610A..88V}. However, in the case of V 0332+53, $E_{\rm cyc}$ in the rising phase of the outburst is larger than in the fading phase. \citet{2016MNRAS.460L..99C} suggest that this decrease in $E_{\rm cyc}$ might be due to a rapid dissipation of the external magnetic field of the NS, affected by diamagnetic screening from the accreted matter. Since we observe an increase of $E_{\rm cyc}$, the scenario proposed by \citet{2016MNRAS.460L..99C} cannot be applied in this case. An alternative interpretation invoked by \citet{2017MNRAS.466.2143D} is that  the energy drift of the CRSF was related to the different emission geometries because of different magnetosphere sizes in the rising and fading phases presumably associated with changes in the internal structure of the accretion disk. This scenario suggests that in comparison to V 0332+53, the geometrical evolution in 1A 0535+262 is opposite: the inner radius of the disc is larger and the polar cap footprint of the magnetic field is smaller during the rising phase compared to those of the fading phase. Consequently, there is a higher accretion column in the supercritical regime of the rising phase. Furthermore, if the observed $E_{\rm cyc}$ in the sub-critical regime is affected by the Doppler effect \citep[see][]{2015MNRAS.454.2539M}, one could expect lower $E_{\rm cyc}$ values in the sub-critical regime of the rising phase. This is because, at a certain luminosity level and a specific height above the polar cup, the local radiation pressure upon the accreting matter is larger for a smaller radius of the hot spot, thereby enhancing the braking effect on the falling material. However, the reasons behind the different inner disk radius evolutions in the two sources remain unclear and are beyond the scope of our current study.

\section{CONCLUSION}
\label{sec:5}
Based on the pulse-to-pulse analysis of Insight-HXMT high cadence observations of 1A 0535+262 during its giant outburst in 2020, we have detected three transitional luminosity levels in the luminosity dependence of the cyclotron line energy ($E_{\rm cyc}$) for the first time in a Type \uppercase\expandafter{\romannumeral2} outburst. 
We find both positive and negative correlations between cyclotron line energy and luminosity at $L_{\rm X}\sim(1$--$3)\times10^{37}\ {\rm erg\ s ^{-1}}$ and $L_{\rm X}\sim(7$--$13)\times10^{37}\ {\rm erg\ s ^{-1}}$, respectively, during both rising and fading phases of the outburst. 
These two luminosity ranges are generally expected to correspond to the typical sub- and super-critical accretion regimes, respectively. 
However, an unexpected plateau in the correlation between the CRSF energy and luminosity is observed in the range of $\sim(3$--$7)\times10^{37}\ {\rm erg\ s^{-1}}$. 
We suggest that the plateau of line energy might result from a combination of Coulomb interactions and radiative pressure decelerating the accreting gas within the accretion column.
Additionally, below the transitional luminosity of $\sim10^{37}\ {\rm erg\ s^{-1}}$, the positive $E_{\rm cyc}$-$L_{\rm X}$ correlation appears to break, and the pulse profile simultaneously transitions to a pure single broad peak.
We propose there are several models might account for the $E_{\rm cyc}$ behavior at $L_{\rm X}\lesssim 10^{37}$ erg s$^{-1}$. 
These findings offer the first comprehensive view of the luminosity dependence of CRSF energy across a broad luminosity range of $\sim(0.03$--$1.3)\times10^{38}\ {\rm erg\ s^{-1}}$, providing valuable insights for further theoretical work.

\section*{Acknowledgements}
The authors are grateful to the anonymous referee for constructive comments that helped us improve this paper. This research has made use of data obtained from the High Energy Astrophysics Science Archive Research Center (HEASARC), provided by NASA’s Goddard Space Flight Center, and the Insight-HXMT mission, a project funded by China National Space Administration (CNSA) and the Chinese Academy of Sciences (CAS). This work is supported by the National Key R\&D Program of China (2021YFA0718500) and the National Natural Science Foundation of China under grants, U1838201, U1838202, 12173103, U2038101 and U1938103. This work is partially supported by International Partnership Program of Chinese Academy of Sciences (Grant No.113111KYSB20190020).

\section*{Data Availability}
The data used in this paper is available from the Insight-HXMT website (\url{http://hxmtweb.ihep.ac.cn/}). Additionally, the high-level data products (e.g. spectra and light curves) underlying this article will be shared upon reasonable request to the lead author.



\bibliographystyle{mnras}
\bibliography{example} 

\begin{thebibliography}{}
\makeatletter
\relax
\def\mn@urlcharsother{\let\do\@makeother \do\$\do\&\do\#\do\^\do\_\do\%\do\~}
\def\mn@doi{\begingroup\mn@urlcharsother \@ifnextchar [ {\mn@doi@} {\mn@doi@[]}}
\def\mn@doi@[#1]#2{\def\@tempa{#1}\ifx\@tempa\@empty \href {http://dx.doi.org/#2} {doi:#2}\else \href {http://dx.doi.org/#2} {#1}\fi \endgroup}
\def\mn@eprint#1#2{\mn@eprint@#1:#2::\@nil}
\def\mn@eprint@arXiv#1{\href {http://arxiv.org/abs/#1} {{\tt arXiv:#1}}}
\def\mn@eprint@dblp#1{\href {http://dblp.uni-trier.de/rec/bibtex/#1.xml} {dblp:#1}}
\def\mn@eprint@#1:#2:#3:#4\@nil{\def\@tempa {#1}\def\@tempb {#2}\def\@tempc {#3}\ifx \@tempc \@empty \let \@tempc \@tempb \let \@tempb \@tempa \fi \ifx \@tempb \@empty \def\@tempb {arXiv}\fi \@ifundefined {mn@eprint@\@tempb}{\@tempb:\@tempc}{\expandafter \expandafter \csname mn@eprint@\@tempb\endcsname \expandafter{\@tempc}}}

\bibitem[\protect\citeauthoryear{{Araya-G{\'o}chez} \& {Harding}}{{Araya-G{\'o}chez} \& {Harding}}{2000}]{2000ApJ...544.1067A}
{Araya-G{\'o}chez} R.~A.,  {Harding} A.~K.,  2000, \mn@doi [\apj] {10.1086/317224}, \href {https://ui.adsabs.harvard.edu/abs/2000ApJ...544.1067A} {544, 1067}

\bibitem[\protect\citeauthoryear{{Arnaud}}{{Arnaud}}{1996}]{1996ASPC..101...17A}
{Arnaud} K.~A.,  1996, in {Jacoby} G.~H.,  {Barnes} J.,  eds,  Astronomical Society of the Pacific Conference Series Vol. 101, Astronomical Data Analysis Software and Systems V. p.~17

\bibitem[\protect\citeauthoryear{{Bailer-Jones}, {Rybizki}, {Fouesneau}, {Mantelet}  \& {Andrae}}{{Bailer-Jones} et~al.}{2018}]{2018AJ....156...58B}
{Bailer-Jones} C.~A.~L.,  {Rybizki} J.,  {Fouesneau} M.,  {Mantelet} G.,   {Andrae} R.,  2018, \mn@doi [\aj] {10.3847/1538-3881/aacb21}, \href {https://ui.adsabs.harvard.edu/abs/2018AJ....156...58B} {156, 58}

\bibitem[\protect\citeauthoryear{{Ballhausen} et~al.,}{{Ballhausen} et~al.}{2017}]{2017A&A...608A.105B}
{Ballhausen} R.,  et~al., 2017, \mn@doi [\aap] {10.1051/0004-6361/201730845}, \href {https://ui.adsabs.harvard.edu/abs/2017A&A...608A.105B} {608, A105}

\bibitem[\protect\citeauthoryear{{Basko} \& {Sunyaev}}{{Basko} \& {Sunyaev}}{1976}]{1976MNRAS.175..395B}
{Basko} M.~M.,  {Sunyaev} R.~A.,  1976, \mn@doi [\mnras] {10.1093/mnras/175.2.395}, \href {https://ui.adsabs.harvard.edu/abs/1976MNRAS.175..395B} {175, 395}

\bibitem[\protect\citeauthoryear{{Becker} et~al.,}{{Becker} et~al.}{2012}]{2012A&A...544A.123B}
{Becker} P.~A.,  et~al., 2012, \mn@doi [\aap] {10.1051/0004-6361/201219065}, \href {https://ui.adsabs.harvard.edu/abs/2012A&A...544A.123B} {544, A123}

\bibitem[\protect\citeauthoryear{{Blum} \& {Kraus}}{{Blum} \& {Kraus}}{2000}]{2000ApJ...529..968B}
{Blum} S.,  {Kraus} U.,  2000, \mn@doi [\apj] {10.1086/308308}, \href {https://ui.adsabs.harvard.edu/abs/2000ApJ...529..968B} {529, 968}

\bibitem[\protect\citeauthoryear{{Burnard}, {Arons}  \& {Klein}}{{Burnard} et~al.}{1991}]{1991ApJ...367..575B}
{Burnard} D.~J.,  {Arons} J.,   {Klein} R.~I.,  1991, \mn@doi [\apj] {10.1086/169653}, \href {https://ui.adsabs.harvard.edu/abs/1991ApJ...367..575B} {367, 575}

\bibitem[\protect\citeauthoryear{{Bykov} \& {Krasil'Shchikov}}{{Bykov} \& {Krasil'Shchikov}}{2004}]{2004AstL...30..309B}
{Bykov} A.~M.,  {Krasil'Shchikov} A.~M.,  2004, \mn@doi [Astronomy Letters] {10.1134/1.1738153}, \href {https://ui.adsabs.harvard.edu/abs/2004AstL...30..309B} {30, 309}

\bibitem[\protect\citeauthoryear{{Caballero} et~al.,}{{Caballero} et~al.}{2007}]{2007A&A...465L..21C}
{Caballero} I.,  et~al., 2007, \mn@doi [\aap] {10.1051/0004-6361:20067032}, \href {https://ui.adsabs.harvard.edu/abs/2007A&A...465L..21C} {465, L21}

\bibitem[\protect\citeauthoryear{{Caballero} et~al.,}{{Caballero} et~al.}{2013}]{2013ApJ...764L..23C}
{Caballero} I.,  et~al., 2013, \mn@doi [\apjl] {10.1088/2041-8205/764/2/L23}, \href {https://ui.adsabs.harvard.edu/abs/2013ApJ...764L..23C} {764, L23}

\bibitem[\protect\citeauthoryear{{Camero-Arranz} et~al.,}{{Camero-Arranz} et~al.}{2012}]{2012ApJ...754...20C}
{Camero-Arranz} A.,  et~al., 2012, \mn@doi [\apj] {10.1088/0004-637X/754/1/20}, \href {https://ui.adsabs.harvard.edu/abs/2012ApJ...754...20C} {754, 20}

\bibitem[\protect\citeauthoryear{{Cao} et~al.,}{{Cao} et~al.}{2020}]{2020SCPMA..6349504C}
{Cao} X.,  et~al., 2020, \mn@doi [Science China Physics, Mechanics, and Astronomy] {10.1007/s11433-019-1506-1}, \href {https://ui.adsabs.harvard.edu/abs/2020SCPMA..6349504C} {63, 249504}

\bibitem[\protect\citeauthoryear{{Chen} et~al.,}{{Chen} et~al.}{2020}]{2020SCPMA..6349505C}
{Chen} Y.,  et~al., 2020, \mn@doi [Science China Physics, Mechanics, and Astronomy] {10.1007/s11433-019-1469-5}, \href {https://ui.adsabs.harvard.edu/abs/2020SCPMA..6349505C} {63, 249505}

\bibitem[\protect\citeauthoryear{{Chen} et~al.,}{{Chen} et~al.}{2021}]{2021ApJ...919...33C}
{Chen} X.,  et~al., 2021, \mn@doi [\apj] {10.3847/1538-4357/ac1268}, \href {https://ui.adsabs.harvard.edu/abs/2021ApJ...919...33C} {919, 33}

\bibitem[\protect\citeauthoryear{{Cusumano}, {La Parola}, {D'A{\`\i}}, {Segreto}, {Tagliaferri}, {Barthelmy}  \& {Gehrels}}{{Cusumano} et~al.}{2016}]{2016MNRAS.460L..99C}
{Cusumano} G.,  {La Parola} V.,  {D'A{\`\i}} A.,  {Segreto} A.,  {Tagliaferri} G.,  {Barthelmy} S.~D.,   {Gehrels} N.,  2016, \mn@doi [\mnras] {10.1093/mnrasl/slw084}, \href {https://ui.adsabs.harvard.edu/abs/2016MNRAS.460L..99C} {460, L99}

\bibitem[\protect\citeauthoryear{{Davidson}}{{Davidson}}{1973}]{1973NPhS..246....1D}
{Davidson} K.,  1973, \mn@doi [Nature Physical Science] {10.1038/physci246001a0}, \href {https://ui.adsabs.harvard.edu/abs/1973NPhS..246....1D} {246, 1}

\bibitem[\protect\citeauthoryear{{Doroshenko}, {Tsygankov}, {Mushtukov}, {Lutovinov}, {Santangelo}, {Suleimanov}  \& {Poutanen}}{{Doroshenko} et~al.}{2017}]{2017MNRAS.466.2143D}
{Doroshenko} V.,  {Tsygankov} S.~S.,  {Mushtukov} A.~A.,  {Lutovinov} A.~A.,  {Santangelo} A.,  {Suleimanov} V.~F.,   {Poutanen} J.,  2017, \mn@doi [\mnras] {10.1093/mnras/stw3236}, \href {https://ui.adsabs.harvard.edu/abs/2017MNRAS.466.2143D} {466, 2143}

\bibitem[\protect\citeauthoryear{{Finger}, {Wilson}  \& {Harmon}}{{Finger} et~al.}{1996}]{1996ApJ...459..288F}
{Finger} M.~H.,  {Wilson} R.~B.,   {Harmon} B.~A.,  1996, \mn@doi [\apj] {10.1086/176892}, \href {https://ui.adsabs.harvard.edu/abs/1996ApJ...459..288F} {459, 288}

\bibitem[\protect\citeauthoryear{{Foreman-Mackey}}{{Foreman-Mackey}}{2016}]{2016JOSS....1...24F}
{Foreman-Mackey} D.,  2016, \mn@doi [The Journal of Open Source Software] {10.21105/joss.00024}, \href {https://ui.adsabs.harvard.edu/abs/2016JOSS....1...24F} {1, 24}

\bibitem[\protect\citeauthoryear{{F{\"u}rst} et~al.,}{{F{\"u}rst} et~al.}{2014}]{2014ApJ...780..133F}
{F{\"u}rst} F.,  et~al., 2014, \mn@doi [\apj] {10.1088/0004-637X/780/2/133}, \href {https://ui.adsabs.harvard.edu/abs/2014ApJ...780..133F} {780, 133}

\bibitem[\protect\citeauthoryear{{Ge} et~al.,}{{Ge} et~al.}{2020}]{2020ApJ...899L..19G}
{Ge} M.~Y.,  et~al., 2020, \mn@doi [\apjl] {10.3847/2041-8213/abac05}, \href {https://ui.adsabs.harvard.edu/abs/2020ApJ...899L..19G} {899, L19}

\bibitem[\protect\citeauthoryear{{Gnedin} \& {Sunyaev}}{{Gnedin} \& {Sunyaev}}{1973}]{1973A&A....25..233G}
{Gnedin} Y.~N.,  {Sunyaev} R.~A.,  1973, \aap, \href {https://ui.adsabs.harvard.edu/abs/1973A&A....25..233G} {25, 233}

\bibitem[\protect\citeauthoryear{{Goodman} \& {Weare}}{{Goodman} \& {Weare}}{2010}]{2010CAMCS...5...65G}
{Goodman} J.,  {Weare} J.,  2010, \mn@doi [Communications in Applied Mathematics and Computational Science] {10.2140/camcos.2010.5.65}, \href {https://ui.adsabs.harvard.edu/abs/2010CAMCS...5...65G} {5, 65}

\bibitem[\protect\citeauthoryear{{Guo} et~al.,}{{Guo} et~al.}{2020}]{2020JHEAp..27...44G}
{Guo} C.-C.,  et~al., 2020, \mn@doi [Journal of High Energy Astrophysics] {10.1016/j.jheap.2020.02.008}, \href {https://ui.adsabs.harvard.edu/abs/2020JHEAp..27...44G} {27, 44}

\bibitem[\protect\citeauthoryear{{Hou}, {Zhang}, {Torres}, {Ji}  \& {Li}}{{Hou} et~al.}{2023}]{Hou2023}
{Hou} X.,  {Zhang} W.,  {Torres} D.~F.,  {Ji} L.,   {Li} J.,  2023, \mn@doi [\apj] {10.3847/1538-4357/acaec7}, \href {https://ui.adsabs.harvard.edu/abs/2023ApJ...944...57H} {944, 57}

\bibitem[\protect\citeauthoryear{{Hu} et~al.,}{{Hu} et~al.}{2023}]{2023ApJ...945..138H}
{Hu} Y.~F.,  et~al., 2023, \mn@doi [\apj] {10.3847/1538-4357/acbc7a}, \href {https://ui.adsabs.harvard.edu/abs/2023ApJ...945..138H} {945, 138}

\bibitem[\protect\citeauthoryear{{Inoue} et~al.,}{{Inoue} et~al.}{2005}]{2005ATel..613....1I}
{Inoue} H.,  et~al., 2005, The Astronomer's Telegram, \href {https://ui.adsabs.harvard.edu/abs/2005ATel..613....1I} {613, 1}

\bibitem[\protect\citeauthoryear{{Isenberg}, {Lamb}  \& {Wang}}{{Isenberg} et~al.}{1998}]{1998ApJ...505..688I}
{Isenberg} M.,  {Lamb} D.~Q.,   {Wang} J. C.~L.,  1998, \mn@doi [\apj] {10.1086/306171}, \href {https://ui.adsabs.harvard.edu/abs/1998ApJ...505..688I} {505, 688}

\bibitem[\protect\citeauthoryear{{Kendziorra} et~al.,}{{Kendziorra} et~al.}{1994}]{1994A&A...291L..31K}
{Kendziorra} E.,  et~al., 1994, \aap, \href {https://ui.adsabs.harvard.edu/abs/1994A&A...291L..31K} {291, L31}

\bibitem[\protect\citeauthoryear{{Klochkov}, {Staubert}, {Santangelo}, {Rothschild}  \& {Ferrigno}}{{Klochkov} et~al.}{2011}]{2011A&A...532A.126K}
{Klochkov} D.,  {Staubert} R.,  {Santangelo} A.,  {Rothschild} R.~E.,   {Ferrigno} C.,  2011, \mn@doi [\aap] {10.1051/0004-6361/201116800}, \href {https://ui.adsabs.harvard.edu/abs/2011A&A...532A.126K} {532, A126}

\bibitem[\protect\citeauthoryear{{Kong} et~al.,}{{Kong} et~al.}{2021}]{2021ApJ...917L..38K}
{Kong} L.~D.,  et~al., 2021, \mn@doi [\apjl] {10.3847/2041-8213/ac1ad3}, \href {https://ui.adsabs.harvard.edu/abs/2021ApJ...917L..38K} {917, L38}

\bibitem[\protect\citeauthoryear{{Kong} et~al.,}{{Kong} et~al.}{2022}]{2022ApJ...932..106K}
{Kong} L.-D.,  et~al., 2022, \mn@doi [\apj] {10.3847/1538-4357/ac6e66}, \href {https://ui.adsabs.harvard.edu/abs/2022ApJ...932..106K} {932, 106}

\bibitem[\protect\citeauthoryear{{Kretschmar} et~al.,}{{Kretschmar} et~al.}{2005}]{2005ATel..601....1K}
{Kretschmar} P.,  et~al., 2005, The Astronomer's Telegram, \href {https://ui.adsabs.harvard.edu/abs/2005ATel..601....1K} {601, 1}

\bibitem[\protect\citeauthoryear{{Krimm} et~al.,}{{Krimm} et~al.}{2013}]{2013ApJS..209...14K}
{Krimm} H.~A.,  et~al., 2013, \mn@doi [\apjs] {10.1088/0067-0049/209/1/14}, \href {https://ui.adsabs.harvard.edu/abs/2013ApJS..209...14K} {209, 14}

\bibitem[\protect\citeauthoryear{{La Parola}, {Cusumano}, {Segreto}  \& {D'A{\`\i}}}{{La Parola} et~al.}{2016}]{2016MNRAS.463..185L}
{La Parola} V.,  {Cusumano} G.,  {Segreto} A.,   {D'A{\`\i}} A.,  2016, \mn@doi [\mnras] {10.1093/mnras/stw1915}, \href {https://ui.adsabs.harvard.edu/abs/2016MNRAS.463..185L} {463, 185}

\bibitem[\protect\citeauthoryear{{Langer} \& {Rappaport}}{{Langer} \& {Rappaport}}{1982}]{1982ApJ...257..733L}
{Langer} S.~H.,  {Rappaport} S.,  1982, \mn@doi [\apj] {10.1086/160028}, \href {https://ui.adsabs.harvard.edu/abs/1982ApJ...257..733L} {257, 733}

\bibitem[\protect\citeauthoryear{{Li} et~al.,}{{Li} et~al.}{2023}]{2023MNRAS.526.3637L}
{Li} P.~P.,  et~al., 2023, \mn@doi [\mnras] {10.1093/mnras/stad2956}, \href {https://ui.adsabs.harvard.edu/abs/2023MNRAS.526.3637L} {526, 3637}

\bibitem[\protect\citeauthoryear{{Liao} et~al.,}{{Liao} et~al.}{2020a}]{2020JHEAp..27...14L}
{Liao} J.-Y.,  et~al., 2020a, \mn@doi [Journal of High Energy Astrophysics] {10.1016/j.jheap.2020.04.002}, \href {https://ui.adsabs.harvard.edu/abs/2020JHEAp..27...14L} {27, 14}

\bibitem[\protect\citeauthoryear{{Liao} et~al.,}{{Liao} et~al.}{2020b}]{2020JHEAp..27...24L}
{Liao} J.-Y.,  et~al., 2020b, \mn@doi [Journal of High Energy Astrophysics] {10.1016/j.jheap.2020.02.010}, \href {https://ui.adsabs.harvard.edu/abs/2020JHEAp..27...24L} {27, 24}

\bibitem[\protect\citeauthoryear{{Liu} et~al.,}{{Liu} et~al.}{2020}]{2020SCPMA..6349503L}
{Liu} C.,  et~al., 2020, \mn@doi [Science China Physics, Mechanics, and Astronomy] {10.1007/s11433-019-1486-x}, \href {https://ui.adsabs.harvard.edu/abs/2020SCPMA..6349503L} {63, 249503}

\bibitem[\protect\citeauthoryear{{Lutovinov}, {Tsygankov}, {Suleimanov}, {Mushtukov}, {Doroshenko}, {Nagirner}  \& {Poutanen}}{{Lutovinov} et~al.}{2015}]{2015MNRAS.448.2175L}
{Lutovinov} A.~A.,  {Tsygankov} S.~S.,  {Suleimanov} V.~F.,  {Mushtukov} A.~A.,  {Doroshenko} V.,  {Nagirner} D.~I.,   {Poutanen} J.,  2015, \mn@doi [\mnras] {10.1093/mnras/stv125}, \href {https://ui.adsabs.harvard.edu/abs/2015MNRAS.448.2175L} {448, 2175}

\bibitem[\protect\citeauthoryear{{Ma} et~al.,}{{Ma} et~al.}{2022}]{2022MNRAS.517.1988M}
{Ma} R.,  et~al., 2022, \mn@doi [\mnras] {10.1093/mnras/stac2768}, \href {https://ui.adsabs.harvard.edu/abs/2022MNRAS.517.1988M} {517, 1988}

\bibitem[\protect\citeauthoryear{{Makishima} et~al.,}{{Makishima} et~al.}{1990}]{1990ApJ...365L..59M}
{Makishima} K.,  et~al., 1990, \mn@doi [\apjl] {10.1086/185888}, \href {https://ui.adsabs.harvard.edu/abs/1990ApJ...365L..59M} {365, L59}

\bibitem[\protect\citeauthoryear{{Malacaria}, {Klochkov}, {Santangelo}  \& {Staubert}}{{Malacaria} et~al.}{2015}]{2015A&A...581A.121M}
{Malacaria} C.,  {Klochkov} D.,  {Santangelo} A.,   {Staubert} R.,  2015, \mn@doi [\aap] {10.1051/0004-6361/201526417}, \href {https://ui.adsabs.harvard.edu/abs/2015A&A...581A.121M} {581, A121}

\bibitem[\protect\citeauthoryear{{Mihara}, {Makishima}  \& {Nagase}}{{Mihara} et~al.}{1998}]{1998AdSpR..22..987M}
{Mihara} T.,  {Makishima} K.,   {Nagase} F.,  1998, \mn@doi [Advances in Space Research] {10.1016/S0273-1177(98)00128-8}, \href {https://ui.adsabs.harvard.edu/abs/1998AdSpR..22..987M} {22, 987}

\bibitem[\protect\citeauthoryear{{Mukherjee}, {Bhattacharya}  \& {Mignone}}{{Mukherjee} et~al.}{2013}]{2013MNRAS.430.1976M}
{Mukherjee} D.,  {Bhattacharya} D.,   {Mignone} A.,  2013, \mn@doi [\mnras] {10.1093/mnras/stt020}, \href {https://ui.adsabs.harvard.edu/abs/2013MNRAS.430.1976M} {430, 1976}

\bibitem[\protect\citeauthoryear{{M{\"u}ller}, {Klochkov}, {Caballero}  \& {Santangelo}}{{M{\"u}ller} et~al.}{2013}]{2013A&A...552A..81M}
{M{\"u}ller} D.,  {Klochkov} D.,  {Caballero} I.,   {Santangelo} A.,  2013, \mn@doi [\aap] {10.1051/0004-6361/201220347}, \href {https://ui.adsabs.harvard.edu/abs/2013A&A...552A..81M} {552, A81}

\bibitem[\protect\citeauthoryear{{Mushtukov} \& {Tsygankov}}{{Mushtukov} \& {Tsygankov}}{2023}]{2023hxga.book..138M}
{Mushtukov} A.,  {Tsygankov} S.,  2023, in , Handbook of X-ray and Gamma-ray Astrophysics (eds. C. Bambi.
p.~138, \mn@doi{10.1007/978-981-16-4544-0_104-1}

\bibitem[\protect\citeauthoryear{{Mushtukov}, {Suleimanov}, {Tsygankov}  \& {Poutanen}}{{Mushtukov} et~al.}{2015a}]{2015MNRAS.447.1847M}
{Mushtukov} A.~A.,  {Suleimanov} V.~F.,  {Tsygankov} S.~S.,   {Poutanen} J.,  2015a, \mn@doi [\mnras] {10.1093/mnras/stu2484}, \href {https://ui.adsabs.harvard.edu/abs/2015MNRAS.447.1847M} {447, 1847}

\bibitem[\protect\citeauthoryear{{Mushtukov}, {Suleimanov}, {Tsygankov}  \& {Poutanen}}{{Mushtukov} et~al.}{2015b}]{2015MNRAS.454.2539M}
{Mushtukov} A.~A.,  {Suleimanov} V.~F.,  {Tsygankov} S.~S.,   {Poutanen} J.,  2015b, \mn@doi [\mnras] {10.1093/mnras/stv2087}, \href {https://ui.adsabs.harvard.edu/abs/2015MNRAS.454.2539M} {454, 2539}

\bibitem[\protect\citeauthoryear{{Mushtukov}, {Tsygankov}, {Serber}, {Suleimanov}  \& {Poutanen}}{{Mushtukov} et~al.}{2015c}]{2015MNRAS.454.2714M}
{Mushtukov} A.~A.,  {Tsygankov} S.~S.,  {Serber} A.~V.,  {Suleimanov} V.~F.,   {Poutanen} J.,  2015c, \mn@doi [\mnras] {10.1093/mnras/stv2182}, \href {https://ui.adsabs.harvard.edu/abs/2015MNRAS.454.2714M} {454, 2714}

\bibitem[\protect\citeauthoryear{{Mushtukov}, {Suleimanov}, {Tsygankov}  \& {Portegies Zwart}}{{Mushtukov} et~al.}{2021}]{2021MNRAS.503.5193M}
{Mushtukov} A.~A.,  {Suleimanov} V.~F.,  {Tsygankov} S.~S.,   {Portegies Zwart} S.,  2021, \mn@doi [\mnras] {10.1093/mnras/stab811}, \href {https://ui.adsabs.harvard.edu/abs/2021MNRAS.503.5193M} {503, 5193}

\bibitem[\protect\citeauthoryear{{Nelson}, {Salpeter}  \& {Wasserman}}{{Nelson} et~al.}{1993}]{1993ApJ...418..874N}
{Nelson} R.~W.,  {Salpeter} E.~E.,   {Wasserman} I.,  1993, \mn@doi [\apj] {10.1086/173445}, \href {https://ui.adsabs.harvard.edu/abs/1993ApJ...418..874N} {418, 874}

\bibitem[\protect\citeauthoryear{{Neumann}, {Avakyan}, {Doroshenko}  \& {Santangelo}}{{Neumann} et~al.}{2023}]{2023A&A...677A.134N}
{Neumann} M.,  {Avakyan} A.,  {Doroshenko} V.,   {Santangelo} A.,  2023, \mn@doi [\aap] {10.1051/0004-6361/202245728}, \href {https://ui.adsabs.harvard.edu/abs/2023A&A...677A.134N} {677, A134}

\bibitem[\protect\citeauthoryear{{Nishimura}}{{Nishimura}}{2014}]{2014ApJ...781...30N}
{Nishimura} O.,  2014, \mn@doi [\apj] {10.1088/0004-637X/781/1/30}, \href {https://ui.adsabs.harvard.edu/abs/2014ApJ...781...30N} {781, 30}

\bibitem[\protect\citeauthoryear{{Poutanen}, {Mushtukov}, {Suleimanov}, {Tsygankov}, {Nagirner}, {Doroshenko}  \& {Lutovinov}}{{Poutanen} et~al.}{2013}]{2013ApJ...777..115P}
{Poutanen} J.,  {Mushtukov} A.~A.,  {Suleimanov} V.~F.,  {Tsygankov} S.~S.,  {Nagirner} D.~I.,  {Doroshenko} V.,   {Lutovinov} A.~A.,  2013, \mn@doi [\apj] {10.1088/0004-637X/777/2/115}, \href {https://ui.adsabs.harvard.edu/abs/2013ApJ...777..115P} {777, 115}

\bibitem[\protect\citeauthoryear{{Reig} \& {Nespoli}}{{Reig} \& {Nespoli}}{2013}]{2013A&A...551A...1R}
{Reig} P.,  {Nespoli} E.,  2013, \mn@doi [\aap] {10.1051/0004-6361/201219806}, \href {https://ui.adsabs.harvard.edu/abs/2013A&A...551A...1R} {551, A1}

\bibitem[\protect\citeauthoryear{{Reig}, {Ma}, {Tao}, {Zhang}, {Zhang}  \& {Doroshenko}}{{Reig} et~al.}{2022}]{2022A&A...659A.178R}
{Reig} P.,  {Ma} R.~C.,  {Tao} L.,  {Zhang} S.,  {Zhang} S.~N.,   {Doroshenko} V.,  2022, \mn@doi [\aap] {10.1051/0004-6361/202142716}, \href {https://ui.adsabs.harvard.edu/abs/2022A&A...659A.178R} {659, A178}

\bibitem[\protect\citeauthoryear{{Rosenberg}, {Eyles}, {Skinner}  \& {Willmore}}{{Rosenberg} et~al.}{1975}]{1975Natur.256..628R}
{Rosenberg} F.~D.,  {Eyles} C.~J.,  {Skinner} G.~K.,   {Willmore} A.~P.,  1975, \mn@doi [\nat] {10.1038/256628a0}, \href {https://ui.adsabs.harvard.edu/abs/1975Natur.256..628R} {256, 628}

\bibitem[\protect\citeauthoryear{{Rothschild} et~al.,}{{Rothschild} et~al.}{2017}]{2017MNRAS.466.2752R}
{Rothschild} R.~E.,  et~al., 2017, \mn@doi [\mnras] {10.1093/mnras/stw3222}, \href {https://ui.adsabs.harvard.edu/abs/2017MNRAS.466.2752R} {466, 2752}

\bibitem[\protect\citeauthoryear{{Sartore}, {Jourdain}  \& {Roques}}{{Sartore} et~al.}{2015}]{2015ApJ...806..193S}
{Sartore} N.,  {Jourdain} E.,   {Roques} J.~P.,  2015, \mn@doi [\apj] {10.1088/0004-637X/806/2/193}, \href {https://ui.adsabs.harvard.edu/abs/2015ApJ...806..193S} {806, 193}

\bibitem[\protect\citeauthoryear{{Sokolova-Lapa} et~al.,}{{Sokolova-Lapa} et~al.}{2021}]{2021A&A...651A..12S}
{Sokolova-Lapa} E.,  et~al., 2021, \mn@doi [\aap] {10.1051/0004-6361/202040228}, \href {https://ui.adsabs.harvard.edu/abs/2021A&A...651A..12S} {651, A12}

\bibitem[\protect\citeauthoryear{{Staubert}, {Shakura}, {Postnov}, {Wilms}, {Rothschild}, {Coburn}, {Rodina}  \& {Klochkov}}{{Staubert} et~al.}{2007}]{2007A&A...465L..25S}
{Staubert} R.,  {Shakura} N.~I.,  {Postnov} K.,  {Wilms} J.,  {Rothschild} R.~E.,  {Coburn} W.,  {Rodina} L.,   {Klochkov} D.,  2007, \mn@doi [\aap] {10.1051/0004-6361:20077098}, \href {https://ui.adsabs.harvard.edu/abs/2007A&A...465L..25S} {465, L25}

\bibitem[\protect\citeauthoryear{{Staubert} et~al.,}{{Staubert} et~al.}{2019}]{2019A&A...622A..61S}
{Staubert} R.,  et~al., 2019, \mn@doi [\aap] {10.1051/0004-6361/201834479}, \href {https://ui.adsabs.harvard.edu/abs/2019A&A...622A..61S} {622, A61}

\bibitem[\protect\citeauthoryear{{Steele}, {Negueruela}, {Coe}  \& {Roche}}{{Steele} et~al.}{1998}]{1998MNRAS.297L...5S}
{Steele} I.~A.,  {Negueruela} I.,  {Coe} M.~J.,   {Roche} P.,  1998, \mn@doi [\mnras] {10.1046/j.1365-8711.1998.01593.x}, \href {https://ui.adsabs.harvard.edu/abs/1998MNRAS.297L...5S} {297, L5}

\bibitem[\protect\citeauthoryear{{Terada} et~al.,}{{Terada} et~al.}{2006}]{2006ApJ...648L.139T}
{Terada} Y.,  et~al., 2006, \mn@doi [\apjl] {10.1086/508018}, \href {https://ui.adsabs.harvard.edu/abs/2006ApJ...648L.139T} {648, L139}

\bibitem[\protect\citeauthoryear{{Tsygankov}, {Lutovinov}, {Churazov}  \& {Sunyaev}}{{Tsygankov} et~al.}{2006}]{2006MNRAS.371...19T}
{Tsygankov} S.~S.,  {Lutovinov} A.~A.,  {Churazov} E.~M.,   {Sunyaev} R.~A.,  2006, \mn@doi [\mnras] {10.1111/j.1365-2966.2006.10610.x}, \href {https://ui.adsabs.harvard.edu/abs/2006MNRAS.371...19T} {371, 19}

\bibitem[\protect\citeauthoryear{{Tsygankov}, {Lutovinov}, {Churazov}  \& {Sunyaev}}{{Tsygankov} et~al.}{2007}]{2007AstL...33..368T}
{Tsygankov} S.~S.,  {Lutovinov} A.~A.,  {Churazov} E.~M.,   {Sunyaev} R.~A.,  2007, \mn@doi [Astronomy Letters] {10.1134/S1063773707060023}, \href {https://ui.adsabs.harvard.edu/abs/2007AstL...33..368T} {33, 368}

\bibitem[\protect\citeauthoryear{{Tsygankov}, {Lutovinov}  \& {Serber}}{{Tsygankov} et~al.}{2010}]{2010MNRAS.401.1628T}
{Tsygankov} S.~S.,  {Lutovinov} A.~A.,   {Serber} A.~V.,  2010, \mn@doi [\mnras] {10.1111/j.1365-2966.2009.15791.x}, \href {https://ui.adsabs.harvard.edu/abs/2010MNRAS.401.1628T} {401, 1628}

\bibitem[\protect\citeauthoryear{{Tsygankov}, {Rouco Escorial}, {Suleimanov}, {Mushtukov}, {Doroshenko}, {Lutovinov}, {Wijnands}  \& {Poutanen}}{{Tsygankov} et~al.}{2019a}]{2019MNRAS.483L.144T}
{Tsygankov} S.~S.,  {Rouco Escorial} A.,  {Suleimanov} V.~F.,  {Mushtukov} A.~A.,  {Doroshenko} V.,  {Lutovinov} A.~A.,  {Wijnands} R.,   {Poutanen} J.,  2019a, \mn@doi [\mnras] {10.1093/mnrasl/sly236}, \href {https://ui.adsabs.harvard.edu/abs/2019MNRAS.483L.144T} {483, L144}

\bibitem[\protect\citeauthoryear{{Tsygankov}, {Doroshenko}, {Mushtukov}, {Suleimanov}, {Lutovinov}  \& {Poutanen}}{{Tsygankov} et~al.}{2019b}]{2019MNRAS.487L..30T}
{Tsygankov} S.~S.,  {Doroshenko} V.,  {Mushtukov} A.~A.,  {Suleimanov} V.~F.,  {Lutovinov} A.~A.,   {Poutanen} J.,  2019b, \mn@doi [\mnras] {10.1093/mnrasl/slz079}, \href {https://ui.adsabs.harvard.edu/abs/2019MNRAS.487L..30T} {487, L30}

\bibitem[\protect\citeauthoryear{{Vybornov}, {Klochkov}, {Gornostaev}, {Postnov}, {Sokolova-Lapa}, {Staubert}, {Pottschmidt}  \& {Santangelo}}{{Vybornov} et~al.}{2017}]{2017A&A...601A.126V}
{Vybornov} V.,  {Klochkov} D.,  {Gornostaev} M.,  {Postnov} K.,  {Sokolova-Lapa} E.,  {Staubert} R.,  {Pottschmidt} K.,   {Santangelo} A.,  2017, \mn@doi [\aap] {10.1051/0004-6361/201630275}, \href {https://ui.adsabs.harvard.edu/abs/2017A&A...601A.126V} {601, A126}

\bibitem[\protect\citeauthoryear{{Vybornov}, {Doroshenko}, {Staubert}  \& {Santangelo}}{{Vybornov} et~al.}{2018}]{2018A&A...610A..88V}
{Vybornov} V.,  {Doroshenko} V.,  {Staubert} R.,   {Santangelo} A.,  2018, \mn@doi [\aap] {10.1051/0004-6361/201731750}, \href {https://ui.adsabs.harvard.edu/abs/2018A&A...610A..88V} {610, A88}

\bibitem[\protect\citeauthoryear{{Walter}, {Lutovinov}, {Bozzo}  \& {Tsygankov}}{{Walter} et~al.}{2015}]{2015A&ARv..23....2W}
{Walter} R.,  {Lutovinov} A.~A.,  {Bozzo} E.,   {Tsygankov} S.~S.,  2015, \mn@doi [\aapr] {10.1007/s00159-015-0082-6}, \href {https://ui.adsabs.harvard.edu/abs/2015A&ARv..23....2W} {23, 2}

\bibitem[\protect\citeauthoryear{{Wang} et~al.,}{{Wang} et~al.}{2022}]{2022ApJ...935..125W}
{Wang} P.~J.,  et~al., 2022, \mn@doi [\apj] {10.3847/1538-4357/ac8230}, \href {https://ui.adsabs.harvard.edu/abs/2022ApJ...935..125W} {935, 125}

\bibitem[\protect\citeauthoryear{{Wilson} \& {Finger}}{{Wilson} \& {Finger}}{2005}]{2005ATel..605....1W}
{Wilson} C.~A.,  {Finger} M.~H.,  2005, The Astronomer's Telegram, \href {https://ui.adsabs.harvard.edu/abs/2005ATel..605....1W} {605, 1}

\bibitem[\protect\citeauthoryear{{Yamamoto}, {Sugizaki}, {Mihara}, {Nakajima}, {Yamaoka}, {Matsuoka}, {Morii}  \& {Makishima}}{{Yamamoto} et~al.}{2011}]{2011PASJ...63S.751Y}
{Yamamoto} T.,  {Sugizaki} M.,  {Mihara} T.,  {Nakajima} M.,  {Yamaoka} K.,  {Matsuoka} M.,  {Morii} M.,   {Makishima} K.,  2011, \mn@doi [\pasj] {10.1093/pasj/63.sp3.S751}, \href {https://ui.adsabs.harvard.edu/abs/2011PASJ...63S.751Y} {63, S751}

\bibitem[\protect\citeauthoryear{{Zel'dovich} \& {Shakura}}{{Zel'dovich} \& {Shakura}}{1969}]{1969SvA....13..175Z}
{Zel'dovich} Y.~B.,  {Shakura} N.~I.,  1969, \sovast, \href {https://ui.adsabs.harvard.edu/abs/1969SvA....13..175Z} {13, 175}

\bibitem[\protect\citeauthoryear{{Zhang}, {Lu}, {Zhang}  \& {Li}}{{Zhang} et~al.}{2014}]{2014SPIE.9144E..21Z}
{Zhang} S.,  {Lu} F.~J.,  {Zhang} S.~N.,   {Li} T.~P.,  2014, in {Takahashi} T.,  {den Herder} J.-W.~A.,   {Bautz} M.,  eds,  Society of Photo-Optical Instrumentation Engineers (SPIE) Conference Series Vol. 9144, Space Telescopes and Instrumentation 2014: Ultraviolet to Gamma Ray. p. 914421, \mn@doi{10.1117/12.2054144}

\bibitem[\protect\citeauthoryear{{Zhang} et~al.,}{{Zhang} et~al.}{2019}]{2019ApJ...879...61Z}
{Zhang} Y.,  et~al., 2019, \mn@doi [\apj] {10.3847/1538-4357/ab22b1}, \href {https://ui.adsabs.harvard.edu/abs/2019ApJ...879...61Z} {879, 61}

\bibitem[\protect\citeauthoryear{{Zhang} et~al.,}{{Zhang} et~al.}{2020}]{2020SCPMA..6349502Z}
{Zhang} S.-N.,  et~al., 2020, \mn@doi [Science China Physics, Mechanics, and Astronomy] {10.1007/s11433-019-1432-6}, \href {https://ui.adsabs.harvard.edu/abs/2020SCPMA..6349502Z} {63, 249502}

\makeatother
\end{thebibliography}




\appendix
\section{Best-fitting spectral parameters}
\begin{landscape}
\begin{table}
\caption{Best-fitting parameters and goodness of fit of all pulse-amplitude-resolved spectra. $E_{\rm cyc}$ and $\tau_1$ are the centroid energy and optical depth of the fundamental cyclotron line, respectively, $\tau_2$ is the optical depth of the harmonic line, $\Gamma$ and $E_{\rm fold}$ are the spectral index and cutoff energy of the \textit{cutoffpl} model, $kT_1$ and $kT_2$ are temperatures of the two \textit{bbodyrad} models, respectively, $E_{\rm Fe}$ is the centroid energy of the iron line, and $L_{\rm X}$ is the luminosity in the 2--150 keV energy range. In our analysis of the low-pulse-amplitude spectrum observed on MJD 59159 we find that the lower-temperature blackbody component is not well constrained. Similarly, the harmonic cyclotron line and the iron emission line in the low-pulse-amplitude spectrum from MJD 59204–59207 observations are also poorly constrained. Therefore, we fixed these parameters to those of the nearest in time spectra where they are well constrained. The uncertainties are reported at a 68\% (1$\sigma$) confidence level.}
\label{tab:A1}
\begin{tabular}{cccccccccccc}
\hline
Time (MJD) & Pulse Bin & $E_{\rm cyc}$ (keV) & $\tau_1$ & $\tau_2$ & $\Gamma$ & $E_{\rm fold}$ (keV) & $kT_1$ & $kT_2$ & $E_{\rm Fe}$ (keV) & $L_{\rm X}$ ($10^{37}$ erg s$^{-1}$) & $\chi^2$ (d.o.f)\\
\hline
59159  & High & $43.47\pm0.54$ & $0.435\pm0.018$ & $0.73\pm0.29$ & $0.427\pm0.023$ & $16.79\pm0.34$ & $1.27\pm0.05$ & $0.616^*$ & $6.36\pm0.15$ & $0.74\pm0.01$ & 299.31 (300) \\ 
  & Low & $45.43\pm0.26$ & $0.451\pm0.011$ & $1.56\pm0.13$ & $0.324\pm0.035$ & $15.61\pm0.22$ & $1.37\pm0.07$ & $0.616\pm0.071$ & $6.25\pm0.09$ & $0.54\pm0.01$ & 313.73 (298) \\ 
59161  & High & $44.41\pm0.19$ & $0.327\pm0.017$ & $0.73\pm0.28$ & $0.338\pm0.023$ & $15.35\pm0.24$ & $1.20\pm0.04$ & $0.080\pm0.042$ & $6.51\pm0.09$ & $1.43\pm0.01$ & 272.06 (298) \\ 
  & Low & $43.86\pm0.50$ & $0.340\pm0.019$ & $1.17\pm0.35$ & $0.357\pm0.022$ & $15.82\pm0.28$ & $1.23\pm0.04$ & $0.078\pm0.042$ & $6.59\pm0.09$ & $1.11\pm0.01$ & 288.74 (298) \\ 
59163--59165 & High & $45.90\pm0.49$ & $0.249\pm0.012$ & $0.77\pm0.18$ & $0.185\pm0.019$ & $14.36\pm0.15$ & $1.45\pm0.04$ & $0.424\pm0.022$ & $6.47\pm0.09$ & $2.34\pm0.01$ & 321.05 (298) \\
  & Low & $45.58\pm0.18$ & $0.266\pm0.012$ & $0.79\pm0.21$ & $0.280\pm0.024$ & $14.89\pm0.21$ & $1.24\pm0.03$ & $0.053\pm0.030$ & $6.50\pm0.07$ & $1.74\pm0.01$ & 290.81 (298) \\  
59167  & High & $45.99\pm0.37$ & $0.140\pm0.008$ & $0.91\pm0.15$ & $-0.295\pm0.013$ & $11.31\pm0.06$ & $1.74\pm0.03$ & $0.546\pm0.011$ & $6.58\pm0.03$ & $7.75\pm0.03$ & 263.67 (298) \\ 
  & Low & $47.02\pm0.30$ & $0.172\pm0.010$ & $0.60\pm0.12$ & $-0.191\pm0.011$ & $11.75\pm0.06$ & $1.67\pm0.03$ & $0.502\pm0.014$ & $6.55\pm0.03$ & $6.30\pm0.02$ & 288.86 (298) \\ 
59168  & High & $45.15\pm0.40$ & $0.145\pm0.007$ & $0.64\pm0.08$ & $-0.352\pm0.012$ & $10.88\pm0.04$ & $1.68\pm0.03$ & $0.539\pm0.013$ & $6.58\pm0.02$ & $9.15\pm0.02$ & 275.54 (298) \\ 
  & Low & $46.25\pm0.39$ & $0.146\pm0.006$ & $0.57\pm0.08$ & $-0.332\pm0.010$ & $11.06\pm0.04$ & $1.74\pm0.02$ & $0.550\pm0.011$ & $6.54\pm0.01$ & $7.80\pm0.02$ & 253.51 (298) \\ 
59169  & High & $43.46\pm0.38$ & $0.123\pm0.007$ & $0.45\pm0.04$ & $-0.399\pm0.013$ & $10.54\pm0.05$ & $1.59\pm0.02$ & $0.522\pm0.008$ & $6.55\pm0.01$ & $10.86\pm0.03$ & 285.87 (298) \\ 
  & Low & $44.83\pm0.23$ & $0.137\pm0.008$ & $0.42\pm0.09$ & $-0.363\pm0.013$ & $10.73\pm0.05$ & $1.62\pm0.02$ & $0.528\pm0.012$ & $6.57\pm0.02$ & $9.30\pm0.03$ & 266.55 (298) \\ 
59170  & High & $42.80\pm0.46$ & $0.143\pm0.008$ & $0.37\pm0.05$ & $-0.474\pm0.011$ & $10.29\pm0.04$ & $1.66\pm0.02$ & $0.554\pm0.009$ & $6.58\pm0.02$ & $11.32\pm0.03$ & 275.71 (298) \\ 
  & Low & $42.99\pm0.28$ & $0.143\pm0.007$ & $0.34\pm0.04$ & $-0.445\pm0.012$ & $10.40\pm0.04$ & $1.61\pm0.02$ & $0.519\pm0.010$ & $6.56\pm0.02$ & $10.22\pm0.02$ & 290.26 (298) \\  
59171  & High & $42.07\pm0.23$ & $0.147\pm0.007$ & $0.27\pm0.05$ & $-0.490\pm0.018$ & $10.21\pm0.06$ & $1.65\pm0.03$ & $0.545\pm0.010$ & $6.61\pm0.02$ & $12.11\pm0.04$ & 260.88 (298) \\ 
  & Low & $42.87\pm0.43$ & $0.137\pm0.008$ & $0.25\pm0.04$ & $-0.460\pm0.010$ & $10.32\pm0.04$ & $1.61\pm0.02$ & $0.530\pm0.008$ & $6.57\pm0.02$ & $10.92\pm0.02$ & 258.64 (298) \\
59172  & High & $43.00\pm0.37$ & $0.133\pm0.007$ & $0.35\pm0.07$ & $-0.439\pm0.011$ & $10.39\pm0.04$ & $1.57\pm0.02$ & $0.516\pm0.009$ & $6.59\pm0.02$ & $11.83\pm0.03$ & 269.46 (298) \\ 
  & Low & $43.49\pm0.35$ & $0.154\pm0.006$ & $0.31\pm0.07$ & $-0.441\pm0.012$ & $10.37\pm0.04$ & $1.64\pm0.02$ & $0.538\pm0.009$ & $6.57\pm0.02$ & $10.61\pm0.03$ & 304.31 (298) \\ 
59173  & High & $43.28\pm0.40$ & $0.147\pm0.010$ & $0.40\pm0.03$ & $-0.450\pm0.016$ & $10.34\pm0.06$ & $1.58\pm0.03$ & $0.535\pm0.013$ & $6.57\pm0.02$ & $12.08\pm0.04$ & 265.77 (298) \\ 
  & Low & $44.34\pm0.38$ & $0.140\pm0.009$ & $0.25\pm0.07$ & $-0.420\pm0.013$ & $10.47\pm0.04$ & $1.57\pm0.02$ & $0.530\pm0.012$ & $6.56\pm0.01$ & $11.05\pm0.03$ & 275.19 (298) \\ 
59174  & High & $43.82\pm0.15$ & $0.138\pm0.008$ & $0.64\pm0.09$ & $-0.450\pm0.013$ & $10.39\pm0.04$ & $1.62\pm0.03$ & $0.536\pm0.009$ & $6.62\pm0.02$ & $11.23\pm0.03$ & 302.66 (298) \\ 
  & Low & $44.87\pm0.33$ & $0.147\pm0.010$ & $0.75\pm0.09$ & $-0.412\pm0.013$ & $10.50\pm0.05$ & $1.53\pm0.02$ & $0.494\pm0.011$ & $6.53\pm0.01$ & $10.32\pm0.03$ & 284.78 (298) \\ 
59175  & High & $43.63\pm0.25$ & $0.141\pm0.009$ & $0.45\pm0.06$ & $-0.444\pm0.018$ & $10.43\pm0.06$ & $1.61\pm0.03$ & $0.554\pm0.011$ & $6.56\pm0.02$ & $11.50\pm0.04$ & 284.79 (298) \\ 
  & Low & $44.12\pm0.27$ & $0.138\pm0.006$ & $0.55\pm0.08$ & $-0.412\pm0.012$ & $10.55\pm0.05$ & $1.53\pm0.02$ & $0.510\pm0.011$ & $6.54\pm0.01$ & $10.32\pm0.03$ & 286.15 (298) \\ 
59177  & High & $44.31\pm0.18$ & $0.129\pm0.009$ & $0.56\pm0.09$ & $-0.407\pm0.013$ & $10.60\pm0.05$ & $1.62\pm0.02$ & $0.552\pm0.011$ & $6.60\pm0.02$ & $10.66\pm0.03$ & 296.86 (298) \\ 
  & Low & $45.22\pm0.48$ & $0.149\pm0.006$ & $0.59\pm0.12$ & $-0.406\pm0.012$ & $10.61\pm0.05$ & $1.60\pm0.02$ & $0.532\pm0.010$ & $6.54\pm0.03$ & $9.81\pm0.03$ & 272.85 (298) \\ 
59178  & High & $43.92\pm0.25$ & $0.155\pm0.007$ & $0.44\pm0.05$ & $-0.442\pm0.011$ & $10.42\pm0.04$ & $1.58\pm0.02$ & $0.531\pm0.009$ & $6.63\pm0.02$ & $10.63\pm0.03$ & 285.33 (298) \\ 
  & Low & $44.91\pm0.37$ & $0.139\pm0.010$ & $0.36\pm0.07$ & $-0.382\pm0.011$ & $10.70\pm0.04$ & $1.57\pm0.02$ & $0.522\pm0.007$ & $6.60\pm0.01$ & $9.55\pm0.03$ & 260.77 (298) \\  
59179  & High & $45.22\pm0.36$ & $0.157\pm0.010$ & $0.59\pm0.09$ & $-0.374\pm0.013$ & $10.68\pm0.04$ & $1.55\pm0.02$ & $0.535\pm0.012$ & $6.56\pm0.02$ & $10.31\pm0.03$ & 276.61 (298) \\
  & Low & $45.23\pm0.42$ & $0.154\pm0.006$ & $0.53\pm0.05$ & $-0.371\pm0.011$ & $10.73\pm0.05$ & $1.51\pm0.02$ & $0.495\pm0.009$ & $6.56\pm0.02$ & $9.19\pm0.02$ & 281.09 (298) \\ 
59180  & High & $45.04\pm0.27$ & $0.156\pm0.006$ & $0.63\pm0.09$ & $-0.393\pm0.012$ & $10.60\pm0.05$ & $1.50\pm0.02$ & $0.499\pm0.009$ & $6.58\pm0.02$ & $10.06\pm0.03$ & 297.68 (298) \\
  & Low & $45.61\pm0.29$ & $0.160\pm0.009$ & $0.74\pm0.08$ & $-0.339\pm0.014$ & $10.91\pm0.06$ & $1.50\pm0.02$ & $0.491\pm0.017$ & $6.60\pm0.02$ & $8.80\pm0.03$ & 270.30 (298) \\  
59181  & High & $45.54\pm0.15$ & $0.172\pm0.008$ & $0.61\pm0.08$ & $-0.405\pm0.009$ & $10.63\pm0.04$ & $1.56\pm0.02$ & $0.509\pm0.009$ & $6.56\pm0.02$ & $9.33\pm0.03$ & 277.17 (298) \\ 
  & Low & $46.31\pm0.35$ & $0.165\pm0.011$ & $0.55\pm0.11$ & $-0.336\pm0.013$ & $10.93\pm0.05$ & $1.51\pm0.02$ & $0.485\pm0.010$ & $6.57\pm0.01$ & $8.34\pm0.03$ & 266.67 (298) \\ 
59182  & High & $45.56\pm0.23$ & $0.152\pm0.008$ & $0.69\pm0.09$ & $-0.340\pm0.011$ & $10.86\pm0.04$ & $1.48\pm0.02$ & $0.479\pm0.015$ & $6.57\pm0.02$ & $8.92\pm0.02$ & 253.54 (298) \\ 
  & Low & $47.13\pm0.44$ & $0.151\pm0.011$ & $1.00\pm0.20$ & $-0.267\pm0.018$ & $11.23\pm0.08$ & $1.46\pm0.02$ & $0.472\pm0.015$ & $6.61\pm0.02$ & $7.75\pm0.03$ & 302.37 (298) \\ 
59183  & High & $46.45\pm0.30$ & $0.155\pm0.009$ & $0.66\pm0.14$ & $-0.297\pm0.012$ & $11.12\pm0.05$ & $1.51\pm0.02$ & $0.494\pm0.013$ & $6.60\pm0.02$ & $8.08\pm0.02$ & 275.71 (298) \\ 
  & Low & $46.72\pm0.37$ & $0.159\pm0.009$ & $0.85\pm0.10$ & $-0.291\pm0.009$ & $11.20\pm0.04$ & $1.50\pm0.01$ & $0.487\pm0.012$ & $6.56\pm0.02$ & $7.13\pm0.02$ & 243.07 (298) \\ 
59184  & High & $46.84\pm0.33$ & $0.169\pm0.005$ & $0.90\pm0.11$ & $-0.275\pm0.007$ & $11.28\pm0.04$ & $1.56\pm0.01$ & $0.470\pm0.009$ & $6.57\pm0.01$ & $7.66\pm0.02$ & 279.04 (298) \\ 
  & Low & $46.63\pm0.35$ & $0.194\pm0.009$ & $0.71\pm0.09$ & $-0.273\pm0.011$ & $11.32\pm0.05$ & $1.63\pm0.02$ & $0.534\pm0.010$ & $6.57\pm0.02$ & $6.60\pm0.02$ & 259.02 (298) \\ 
\end{tabular}
\end{table}
\end{landscape}

\begin{landscape}
\begin{table}
\contcaption{}
\begin{tabular}{cccccccccccc}
\hline
Time (MJD) & Pulse Bin & $E_{\rm cyc}$ (keV) & $\tau_1$ & $\tau_2$ & $\Gamma$ & $E_{\rm fold}$ (keV) & $kT_1$ & $kT_2$ & $E_{\rm Fe}$ (keV) & $L_{\rm X}$ ($10^{37}$ erg s$^{-1}$) & $\chi^2$ (d.o.f)\\
\hline
59185  & High & $46.95\pm0.35$ & $0.168\pm0.006$ & $0.89\pm0.15$ & $-0.260\pm0.014$ & $11.29\pm0.06$ & $1.55\pm0.02$ & $0.493\pm0.010$ & $6.59\pm0.02$ & $7.40\pm0.02$ & 282.54 (298) \\ 
  & Low & $48.26\pm0.39$ & $0.169\pm0.007$ & $1.04\pm0.08$ & $-0.199\pm0.010$ & $11.61\pm0.05$ & $1.52\pm0.02$ & $0.474\pm0.010$ & $6.59\pm0.01$ & $6.61\pm0.02$ & 274.05 (298) \\  
59186  & High & $47.99\pm0.37$ & $0.179\pm0.009$ & $0.88\pm0.13$ & $-0.224\pm0.012$ & $11.57\pm0.06$ & $1.55\pm0.02$ & $0.489\pm0.012$ & $6.58\pm0.02$ & $6.78\pm0.02$ & 296.02 (298) \\ 
  & Low & $47.77\pm0.41$ & $0.183\pm0.010$ & $0.74\pm0.14$ & $-0.215\pm0.015$ & $11.64\pm0.07$ & $1.53\pm0.02$ & $0.460\pm0.013$ & $6.55\pm0.02$ & $6.03\pm0.02$ & 274.86 (298) \\ 
59187  & High & $46.84\pm0.28$ & $0.169\pm0.010$ & $0.82\pm0.12$ & $-0.205\pm0.012$ & $11.71\pm0.05$ & $1.55\pm0.02$ & $0.491\pm0.016$ & $6.59\pm0.01$ & $6.48\pm0.02$ & 291.40 (298) \\ 
  & Low & $48.02\pm0.46$ & $0.200\pm0.006$ & $1.04\pm0.11$ & $-0.177\pm0.011$ & $11.96\pm0.06$ & $1.59\pm0.02$ & $0.482\pm0.011$ & $6.57\pm0.03$ & $5.59\pm0.01$ & 296.40 (298) \\ 
59188  & High & $47.40\pm0.39$ & $0.193\pm0.010$ & $1.29\pm0.19$ & $-0.149\pm0.014$ & $12.00\pm0.07$ & $1.48\pm0.02$ & $0.443\pm0.016$ & $6.56\pm0.02$ & $5.60\pm0.02$ & 300.27 (298) \\ 
  & Low & $47.92\pm0.38$ & $0.198\pm0.011$ & $1.18\pm0.14$ & $-0.126\pm0.016$ & $12.26\pm0.08$ & $1.52\pm0.02$ & $0.479\pm0.018$ & $6.62\pm0.03$ & $4.90\pm0.02$ & 291.29 (298) \\ 
59189  & High & $47.99\pm0.47$ & $0.167\pm0.009$ & $1.97\pm0.30$ & $-0.126\pm0.013$ & $12.07\pm0.06$ & $1.52\pm0.03$ & $0.480\pm0.018$ & $6.60\pm0.03$ & $5.27\pm0.02$ & 291.40 (298) \\ 
  & Low & $47.96\pm0.51$ & $0.170\pm0.009$ & $2.35\pm0.40$ & $-0.109\pm0.007$ & $12.14\pm0.05$ & $1.52\pm0.02$ & $0.428\pm0.012$ & $6.58\pm0.02$ & $4.65\pm0.02$ & 296.40 (298) \\ 
59190  & High & $47.14\pm0.40$ & $0.187\pm0.008$ & $1.05\pm0.14$ & $-0.079\pm0.016$ & $12.43\pm0.09$ & $1.55\pm0.02$ & $0.469\pm0.015$ & $6.62\pm0.02$ & $5.11\pm0.02$ & 305.41 (298) \\ 
  & Low & $48.37\pm0.33$ & $0.201\pm0.009$ & $1.53\pm0.18$ & $-0.002\pm0.012$ & $12.81\pm0.08$ & $1.43\pm0.01$ & $0.394\pm0.013$ & $6.55\pm0.02$ & $4.45\pm0.01$ & 286.76 (298) \\ 
59191  & High & $48.05\pm0.43$ & $0.190\pm0.011$ & $1.25\pm0.13$ & $0.028\pm0.011$ & $12.98\pm0.07$ & $1.47\pm0.02$ & $0.411\pm0.016$ & $6.59\pm0.03$ & $4.32\pm0.02$ & 281.94 (298) \\ 
  & Low & $47.38\pm0.36$ & $0.213\pm0.010$ & $1.29\pm0.20$ & $0.047\pm0.012$ & $13.14\pm0.07$ & $1.44\pm0.01$ & $0.349\pm0.011$ & $6.59\pm0.02$ & $3.67\pm0.01$ & 287.57 (298) \\ 
59192  & High & $47.80\pm0.35$ & $0.203\pm0.006$ & $1.20\pm0.17$ & $0.083\pm0.013$ & $13.32\pm0.09$ & $1.44\pm0.02$ & $0.382\pm0.012$ & $6.66\pm0.04$ & $3.70\pm0.01$ & 291.98 (298) \\ 
  & Low & $48.14\pm0.44$ & $0.224\pm0.008$ & $1.10\pm0.28$ & $0.124\pm0.014$ & $13.68\pm0.08$ & $1.46\pm0.02$ & $0.336\pm0.014$ & $6.61\pm0.02$ & $3.26\pm0.01$ & 288.68 (298) \\ 
59193  & High & $47.80\pm0.35$ & $0.203\pm0.006$ & $1.20\pm0.17$ & $0.083\pm0.013$ & $13.32\pm0.09$ & $1.44\pm0.02$ & $0.382\pm0.012$ & $6.66\pm0.04$ & $3.70\pm0.01$ & 274.93 (298) \\ 
  & Low & $47.97\pm0.47$ & $0.227\pm0.012$ & $0.89\pm0.14$ & $0.160\pm0.016$ & $13.82\pm0.11$ & $1.42\pm0.01$ & $0.322\pm0.012$ & $6.60\pm0.03$ & $3.11\pm0.01$ & 274.61 (298) \\  
59194  & High & $47.71\pm0.32$ & $0.224\pm0.007$ & $1.17\pm0.17$ & $0.158\pm0.017$ & $13.89\pm0.13$ & $1.41\pm0.01$ & $0.331\pm0.013$ & $6.55\pm0.02$ & $3.27\pm0.01$ & 269.14 (298) \\ 
  & Low & $47.71\pm0.32$ & $0.224\pm0.007$ & $1.17\pm0.17$ & $0.158\pm0.017$ & $13.89\pm0.13$ & $1.41\pm0.01$ & $0.331\pm0.013$ & $6.55\pm0.02$ & $3.27\pm0.01$ & 284.69 (298) \\
59195  & High & $48.25\pm0.24$ & $0.223\pm0.012$ & $1.38\pm0.15$ & $0.240\pm0.013$ & $14.33\pm0.10$ & $1.39\pm0.02$ & $0.315\pm0.010$ & $6.61\pm0.03$ & $2.83\pm0.01$ & 319.43 (298) \\ 
  & Low & $47.26\pm0.44$ & $0.246\pm0.013$ & $0.60\pm0.18$ & $0.223\pm0.011$ & $14.21\pm0.09$ & $1.40\pm0.01$ & $0.325\pm0.013$ & $6.59\pm0.02$ & $2.47\pm0.01$ & 277.37 (298) \\ 
59196  & High & $47.31\pm0.37$ & $0.270\pm0.008$ & $0.87\pm0.15$ & $0.215\pm0.013$ & $14.21\pm0.13$ & $1.40\pm0.02$ & $0.359\pm0.013$ & $6.54\pm0.03$ & $2.54\pm0.01$ & 289.88 (298) \\ 
  & Low & $47.22\pm0.48$ & $0.284\pm0.013$ & $1.05\pm0.23$ & $0.306\pm0.015$ & $14.88\pm0.10$ & $1.38\pm0.02$ & $0.318\pm0.015$ & $6.64\pm0.03$ & $2.14\pm0.01$ & 269.65 (298) \\ 
59197  & High & $47.74\pm0.40$ & $0.286\pm0.014$ & $0.98\pm0.21$ & $0.294\pm0.015$ & $15.11\pm0.15$ & $1.37\pm0.02$ & $0.307\pm0.015$ & $6.61\pm0.03$ & $2.06\pm0.01$ & 278.44 (298) \\ 
  & Low & $45.98\pm0.37$ & $0.304\pm0.016$ & $0.96\pm0.24$ & $0.370\pm0.018$ & $15.73\pm0.17$ & $1.38\pm0.01$ & $0.242\pm0.012$ & $6.66\pm0.03$ & $1.67\pm0.01$ & 292.85 (298) \\ 
59198  & High & $46.67\pm0.35$ & $0.292\pm0.014$ & $0.77\pm0.19$ & $0.293\pm0.015$ & $14.77\pm0.12$ & $1.39\pm0.02$ & $0.337\pm0.019$ & $6.62\pm0.02$ & $1.76\pm0.01$ & 313.35 (298) \\ 
  & Low & $46.37\pm0.48$ & $0.319\pm0.009$ & $0.27\pm0.15$ & $0.324\pm0.017$ & $15.16\pm0.16$ & $1.55\pm0.04$ & $0.531\pm0.033$ & $6.44\pm0.06$ & $1.51\pm0.01$ & 298.55 (298) \\ 
59199  & High & $46.23\pm0.33$ & $0.334\pm0.015$ & $0.45\pm0.25$ & $0.301\pm0.019$ & $14.97\pm0.17$ & $1.53\pm0.03$ & $0.368\pm0.023$ & $6.43\pm0.09$ & $1.49\pm0.01$ & 283.89 (298) \\ 
  & Low & $45.81\pm0.40$ & $0.376\pm0.016$ & $0.70\pm0.26$ & $0.368\pm0.021$ & $15.88\pm0.19$ & $1.57\pm0.05$ & $0.462\pm0.027$ & $6.56\pm0.12$ & $1.23\pm0.01$ & 287.59 (298) \\ 
59200  & High & $46.11\pm0.36$ & $0.352\pm0.017$ & $0.50\pm0.25$ & $0.377\pm0.020$ & $15.97\pm0.16$ & $1.63\pm0.04$ & $0.367\pm0.021$ & $6.69\pm0.09$ & $1.26\pm0.01$ & 285.84 (298) \\ 
  & Low & $46.58\pm0.55$ & $0.389\pm0.011$ & $1.11\pm0.37$ & $0.373\pm0.029$ & $15.84\pm0.28$ & $1.58\pm0.06$ & $0.451\pm0.044$ & $6.62\pm0.09$ & $1.03\pm0.01$ & 275.07 (298) \\ 
59201  & High & $46.36\pm0.51$ & $0.349\pm0.016$ & $0.27\pm0.16$ & $0.413\pm0.018$ & $16.00\pm0.21$ & $1.45\pm0.04$ & $0.379\pm0.036$ & $6.58\pm0.08$ & $1.03\pm0.01$ & 292.08 (298) \\ 
  & Low & $46.15\pm0.49$ & $0.405\pm0.014$ & $1.42\pm0.18$ & $0.360\pm0.029$ & $16.18\pm0.27$ & $1.72\pm0.05$ & $0.470\pm0.031$ & $6.48\pm0.25$ & $0.85\pm0.01$ & 306.11 (298) \\ 
59202  & High & $46.72\pm0.41$ & $0.415\pm0.016$ & $1.04\pm0.23$ & $0.334\pm0.026$ & $15.71\pm0.25$ & $1.70\pm0.05$ & $0.628\pm0.031$ & $6.50\pm0.11$ & $0.78\pm0.01$ & 286.04 (298) \\ 
  & Low & $47.37\pm0.47$ & $0.445\pm0.023$ & $1.58\pm0.48$ & $0.528\pm0.026$ & $16.95\pm0.36$ & $1.48\pm0.06$ & $0.437\pm0.028$ & $6.36\pm0.09$ & $0.55\pm0.01$ & 280.25 (298) \\ 
59204--59207  & High & $47.84\pm1.00$ & $0.489\pm0.028$ & $2.81\pm1.44$ & $0.472\pm0.037$ & $17.31\pm0.55$ & $1.88\pm0.12$ & $0.741\pm0.054$ & $6.62\pm0.20$ & $0.42\pm0.01$ & 281.65 (298) \\ 
  & Low & $49.78\pm1.66$ & $0.359\pm0.081$ & $2.81^*$ & $0.702\pm0.042$ & $18.35\pm0.76$ & $1.32\pm0.08$ & $0.239\pm0.128$ & $6.62^*$ & $0.30\pm0.01$ & 276.45 (301) \\
  \hline
\end{tabular}
\end{table}
\end{landscape}

\section{MCMC Parameter Probability Distributions}
\begin{figure*}
    \includegraphics[width=\textwidth]{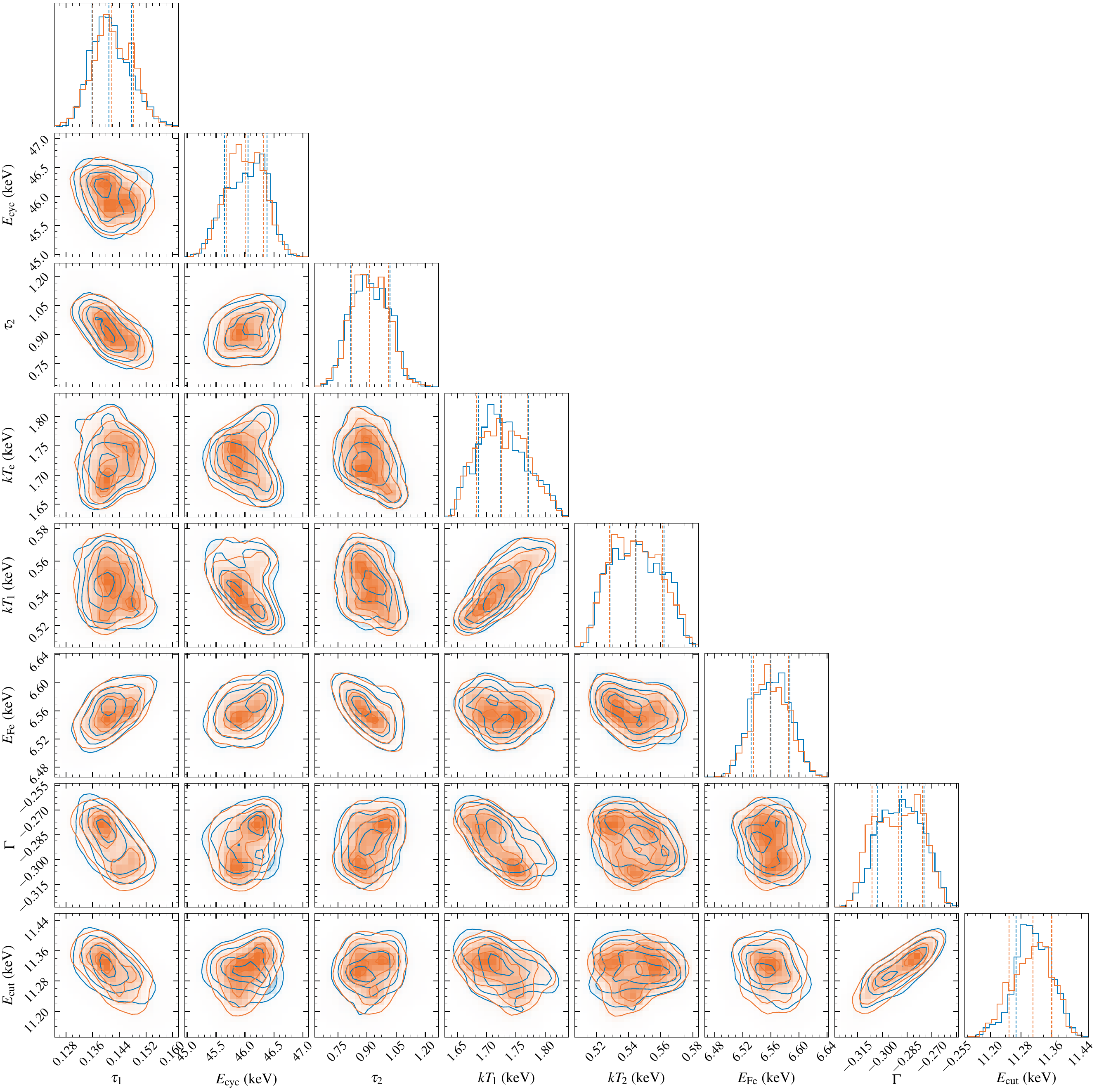}
    \caption{One- and two-dimensional projections of the posterior probability distributions, and the 0.16, 0.5, and 0.84 quantile contours derived from the MCMC analysis for each free physical parameter. To test the convergence, we compare the one- and two-dimensional projections of the posterior distributions from the first (blue) and second (orange) halves of the chain, and we find no large differences. This illustration corresponds to the spectral fitting of the high-pulse-amplitude bin on MJD 59167. The best-fitting parameters of this example are summarized in Table~\ref{tab:1}.}
    \label{fig:mcmc}
\end{figure*}

In this appendix, we show an example (the high-pulse-amplitude spectrum on MJD 59167) of the MCMC analysis for the spectral fitting model presented in Sec.~\ref{sec:3}. The best-fitting parameters of this example spectral fitting are summarized in Table~\ref{tab:1}. We use the Goodman-Weare algorithm with 8 walkers and a total length of 40,000 to perform the MCMC analysis, and the initial 2000 elements are discarded as the burn-in period during which the chain reaches its stationary state. We find that the autocorrelation length is typically 1000 elements, so the net number of independent samples in the parameter space we have is on the order of $10^4$. Furthermore, we compare the one- and two-dimensional projections of the posterior distributions for each parameter from the first and second halves of the chain to test the convergence, and we find no significant differences (see Fig.~\ref{fig:mcmc}). Specifically, the difference of the parameter estimation between posteriors in the halves of the chain is <1\% for each parameter in this representative example. The contour maps and probability distributions are plotted using the corner package \citep{2016JOSS....1...24F}.

\bsp	
\label{lastpage}
\end{document}